\documentclass[prx, aps, twocolumn, superscriptaddress, reprint, nobalancelastpage, nofootinbib, noeprint]{revtex4-2}

\usepackage[pdftex]{graphicx}
\usepackage{hyperref}
\usepackage{amsmath}
\usepackage{physics}
\usepackage{comment}
\usepackage{xcolor}
\usepackage{bbold}

\usepackage{glossaries}
\usepackage{qcircuit}
\usepackage{cleveref}

\newcommand{\MS}{{M{\o}lmer-S{\o}rensen }}

\begin{document}

\begin{abstract}
    The performance of quantum gate operations is experimentally determined by how correct operational parameters can be determined and set, and how stable these parameters can be maintained. In addition, gates acting on different sets of qubits require unique sets of control parameters. Thus, an efficient multi-dimensional parameter estimation procedure is crucial to calibrate even medium sized quantum processors. Here, we develop and characterize an efficient calibration protocol to automatically estimate and adjust experimental parameters of the widely used \MS entangling gate operation in a trapped ion quantum information processor. The protocol exploits Bayesian parameter estimation methods which includes a stopping criterion based on a desired gate infidelity. We experimentally demonstrate a median gate infidelity of $1.3(1)\cdot 10^{-3}$, requiring only $1200\pm500$ experimental cycles, while completing the entire gate calibration procedure in less than one minute.  This approach is applicable to other quantum information processor architectures with known or  sufficiently characterized theoretical models. 
\end{abstract}

\title{Experimental Bayesian calibration of trapped ion entangling operations}

\author{Lukas Gerster}
\email{lukas.gerster@uibk.ac.at}
\affiliation{Institut f\"ur Experimentalphysik, Universit\"at Innsbruck, Technikerstraße 25/4, 6020 Innsbruck, Austria}
\author{Fernando Martínez-García}
\affiliation{Department of Physics, Swansea University, Singleton Park, Swansea SA2 8PP, United Kingdom}
\author{Pavel Hrmo} 
\email{present address: Trapped Ion Quantum Information Group, Institute for Quantum Electronics, ETH Zurich, 8093 Zurich, Switzerland}
\author{Martin van Mourik}
\author{Benjamin Wilhelm}
\affiliation{Institut f\"ur Experimentalphysik, Universit\"at Innsbruck, Technikerstraße 25/4, 6020 Innsbruck, Austria}
\author{Davide Vodola}
\affiliation{Dipartimento di Fisica e Astronomia Augusto Righi, Via Irnerio 46, Bologna, Italy}
\author{Markus Müller}
\affiliation{Institute for Quantum Information, RWTH Aachen University, D-52056 Aachen, Germany}
\affiliation{Peter Gr\"unberg Institute, Theoretical Nanoelectronics, Forschungszentrum J\"ulich, D-52425 J\"ulich, Germany}
\author{Rainer Blatt}
\affiliation{Institut f\"ur Experimentalphysik, Universit\"at Innsbruck, Technikerstraße 25/4, 6020 Innsbruck, Austria}
\affiliation{Institut für Quantenoptik und Quanteninformation,
Österreichische Akademie der Wissenschaften, Technikerstraße 21a, 6020 Innsbruck, Austria}
\author{Philipp Schindler}
\affiliation{Institut f\"ur Experimentalphysik, Universit\"at Innsbruck, Technikerstraße 25/4, 6020 Innsbruck, Austria}
\author{Thomas Monz}
\affiliation{Institut f\"ur Experimentalphysik, Universit\"at Innsbruck, Technikerstraße 25/4, 6020 Innsbruck, Austria}
\affiliation{AQT, Technikerstraße 17, 6020 Innsbruck, Austria}

\date{\today}

\maketitle

\section{Introduction}

The development of quantum information processors has made rapid progress in recent years. The leading paradigm for quantum computation is the circuit model where local operations and two- or multi-qubit entangling operations provide a gate set that allows for universal application of quantum circuits. New technological and theoretical developments in various quantum computing platforms \cite{Saffman2016,Wendin2017,McArdle2020,Huang2020,Slussarenko2019,Bruzewicz2019a} have pushed the fidelities of the available entangling gate operations closer to the parameter regime needed for fault-tolerant quantum error correction \cite{Ballance2016,Hughes2020,Wang2020,Leung2018,Gaebler2016,Hong2020,Barends2014,Rol2019,Huang2019,Graham2019}. Furthermore, even without quantum error correction there is the expectation that with moderately sized systems with upwards of about 50 physical qubits with sufficiently high fidelities quantum advantage can be observed \cite{Preskill2018,Montanaro2016,Bharti2021}. However, achieving the necessary fidelities requires a precise calibration of the various classical experimental control parameters that determine the realized Hamiltonians that generate single-qubit and in particular two- or multi-qubit entangling gate operations. Typically such calibrations need to be performed by a highly trained operator who is familiar with the system. Furthermore, the control parameters are liable to drifts and will require some form of feedback to maintain the desired fidelity during the course of operation. 
It is thus highly desirable for a quantum computing platform to implement an automation procedure that can determine the optimal control parameters accurately, such that it can be operated by an end user with the option for periodic re-calibration \cite{Patterson2019,Kelly2016,Arute2019, Klimov2020}. As quantum computing platforms mature and move to remotely-accessed services, such automated calibration routines will become indispensable to keep the machines at peak performance without the need for in-person maintenance, allowing the end user to focus on the algorithmic applications rather than calibration of the hardware. Ultimately, as the complexity of the control system will grow with the size of the qubit register, manual calibration of all couplings will no longer be feasible, and automation routines will underpin reliable long-term operation of the system.

At first sight, the problem of calibrating multiple control parameters would not appear difficult if their action on the quantum system could be independently measured and the parameter corrected accordingly. For example, Ramsey spectroscopy in both frequentist \cite{Li2018,Akerman2015,Schirmer2015} and Bayesian \cite{Martinez-Garcia2019,Li2018,Teklu2009,Yang2018,Schirmer2015} form can be used to determine the mismatch between a qubit transition frequency and the driving field. Indeed, combined with a Rabi frequency measurement \cite{Kiilerich2015} to determine the applied field strength, single-qubit operations can be efficiently calibrated and traced in time using the minimum number of experimental measurements to correct for drifts \cite{Proctor2020,Ralph2011}. However, two-qubit entangling gates often require a more complex combination of driving fields that can have combined effects, which can not simply be measured individually without assessing the gate performance itself.

As a specific example, we will consider the M{\o}lmer-S{\o}rensen (MS) gate \cite{Sorensen1999, sorensen2000entanglement}, which is one of the leading implementations of entangling operations in trapped ion systems. In the following we will focus on the case of applying the MS gate on two qubits only, as circuits constructed from two-qubit gates rather than multi-qubit gate operations are the most widely pursued approach to fulfill fault-tolerant circuit design properties. The MS gate requires a bichromatic driving field whose two frequency components are symmetrically detuned from the qubit transition of the two ions, naively yielding four independent control parameters (two frequencies and two intensities). However, because the fields are applied simultaneously, the total dynamical AC Stark shift arising from a multi-level atom needs to be compensated using either the frequency or intensity of both fields \cite{Kirchmair2009}, and the relative phase of the fields starts to play a role \cite{Roos2008}.
This leads to these parameters being non-linearly correlated and thus suggests that their calibration be carried out directly by measuring the gate action on a known input state rather than independently estimating each parameter. 
Such measurements are routinely used in the ``manual" parameter optimisation, whereby the experimentalist has prior knowledge of the expected outcome of an imperfectly calibrated gate, and uses a scheme to iteratively measure and improve these outcomes while changing the experimental control parameters. In the context of the MS gate, this requires knowledge of the multi-dimensional parameter landscape describing the probability of inducing spin flips, i.e.~changes in the electronic population, on the two ionic qubits as a function of the various control parameters. 
To evaluate the system response to the control parameters we require an efficient characterisation of the gate action. While the gold standard for such a procedure is process tomography \cite{Riebe2006}, it is not without flaws in that it is inherently sensitive to state preparation and measurement (SPAM) errors, is prohibitively slow since the number of required measurements scales exponentially with the number of qubits \cite{Chuang1997}, and can also be problematic when faced with systematic errors \cite{Merkel2013}. 
Therefore, recently, a number of alternative techniques have been developed that allow for either faster or more rigorous characterisation of the gate performance, using randomized benchmarking \cite{Knill2008,Mavadia2017}, cycle benchmarking \cite{Erhard2019}, gate set tomography \cite{Blume-Kohout2017,Mavadia2017} and adaptive methods based on Bayesian estimation \cite{Pogorelov2017,Granade2017}.
However, in practice a full characterization of the gate relative to all experimental control parameters is not required to calibrate the gate. An often used practical method to enhance sensitivity to miscalibrated gate parameters is to instead apply a sequence of concatenated identical gates to a single input state and compare the measured electronic populations (spin excitations) to the expected values of the output states, e.g.~of Bell states in the calibration of two-qubit gates. The trade-off is that this excitation landscape becomes increasingly complicated in the number $N_g$ of gates, with many local minima, and thus requires a judicious choice of $N_g$ according to the uncertainty on the control parameters. With a large amount of initial uncertainty calibration begins with $N_g=1$, but as this uncertainty diminishes, larger $N_g$ values can be used to increase the calibration precision.

In this manuscript we will demonstrate an approach to automatize this calibration process using a Bayesian estimation technique to simultaneously determine and optimize key control parameters of the two-qubit MS gate Hamiltonian. The motivation behind selecting this approach over gradient descent least squares \cite{Rol2017} or machine learning methods \cite{Greplova2017} is that it aims to reduce the required number of measurements to obtain an accurate parameter estimate while retaining the learnt information when switching between measurements using a different number of gates $N_g$. Importantly, the Bayesian approach presented here also provides an intrinsic measure of the uncertainty of the estimate, informing us about the progress of the estimation and thereby a quantitative criterion for when to stop the optimization routine once the optimization target has been reached. A requirement for such a Bayesian protocol to work is a precise knowledge of the underlying theoretical model and dominant imperfections. In our case, the Hamiltonian governing the MS gate operation depends on four control parameters, which we introduce and discuss in detail below: \textit{Sideband detuning}, \textit{center line detuning}, \textit{Rabi frequency} and \textit{phase difference}. The action of this Hamiltonian and the associated noise processes are well understood and map accurately to experimental data as we will show in Sec.~\ref{sec:model_comparison}. 

The manuscript is structured as follows. In Sec.~\ref{sec:ModelMS} we review the underlying model of the \MS gate and introduce the experimentally relevant control parameters that need to be calibrated. In
Sec.~\ref{sec:BayesInf} we show how we can iteratively estimate the control parameters using a Bayesian protocol. In Sec.~\ref{sec:AutoAlgo} we describe the strategies for the choice of measurement setting, introduce the termination criterion of the optimisation routine and experimentally evaluate the performance of the algorithm. Sec.~\ref{sec:conOut} provides conclusions and an outlook. 

\begin{figure*}[ht]
    \centering
    \includegraphics[width=0.9\textwidth]{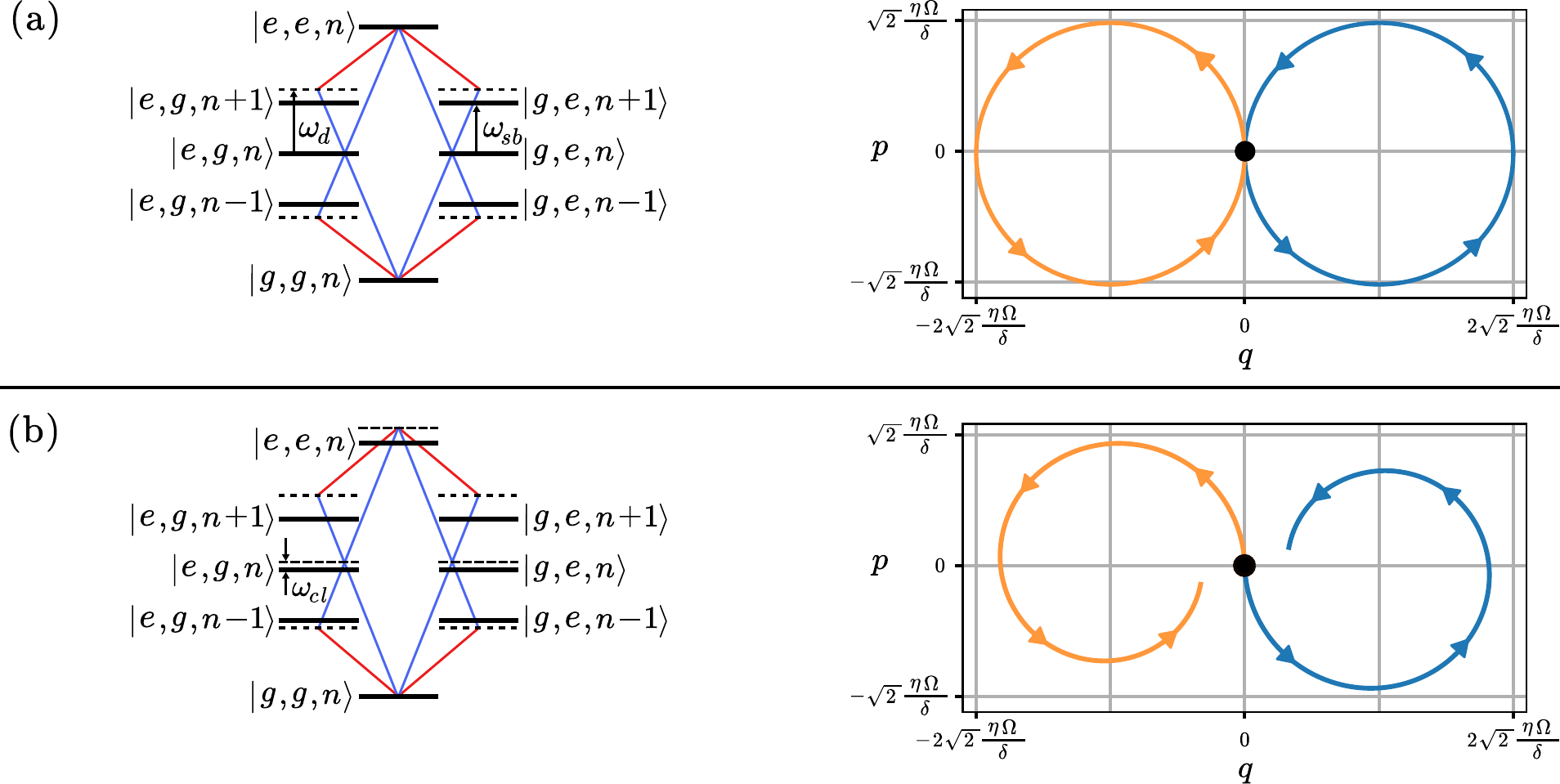}
    \caption{(a) Left: Energy level diagram for two ions with quantized center of mass vibrational mode of frequency $\omega_{sb}$ interacting with a bichromatic laser of frequencies $\omega_1=\omega_{eg}+\omega_d$ and $\omega_2=\omega_{eg}-\omega_d$. The laser allows a resonant two-photon transition between $\ket{g,g,n}$ and $\ket{e,e,n}$. There are four different paths, each of them going through an intermediate virtual state separated by $\delta=\omega_{sb}-\omega_d$ from the first motional sidebands ($\ket{g, e, n \pm 1}, \ket{e, g, n \pm 1}$). A similar diagram can be drawn for the two-photon transition between the states $\ket{e,g,n}$ and $\ket{g,e,n}$. Right: Trajectory in phase space induced by the gate with constant Rabi frequency with the ions in the $+1$ eigenstate of $S_\varphi$ (in orange) and the $-1$ eigenstate (in blue), with the black circle indicating the initial motional state. If we apply the gate for the correct amount of time the trajectory returns to the initial point in phase space. The rotation introduced by the gate is given by the area enclosed by the trajectory. (b) Left: A source of error in the MS gate is when the frequencies of the laser are not centered around the carrier, but instead have a center line detuning $\omega_{cl}$, which we consider to be time-independent in this figure. 
    The bigger this frequency $\omega_{cl}$ is, the less resonant the two-photon process is, which decreases the fidelity of the gate. Right: Similar phase space plot as in the previous case but with $\omega_{cl}\neq 0$. The trajectories for this case are dependent on the initial motional state, where for this figure we have considered it to be $\ket{n=0}$.}
\label{fig:MS_levels}
\end{figure*}

\section{Experimental setup and M{\o}lmer-S{\o}rensen Gate}
\label{sec:ModelMS}

In this section we describe the trapped ion setup that we use to generate entangling operations. We briefly review the physics of the MS gate and introduce the main control parameters that need to be calibrated. 

\subsection{Experimental Setup}

The experiments in this manuscript are performed on $^{40}\mathrm{Ca}^+$ ions confined in a microstructured radio-frequency ion surface trap \cite{brandl2016cryogenic}. Qubits are encoded in the computational subspace formed by the $4^2S_{1/2,-1/2}\equiv \ket{g}$ electronic ground state and the metastable excited  $3^2D_{5/2,-1/2}\equiv \ket{e}$ state. The ions form a crystal whose dynamics is described in terms of one-dimensional coupled harmonic oscillators. These dynamics can then be decomposed into normal modes. Without loss of generality, we consider only the lowest frequency mode corresponding to the center-of-mass (COM) motion in our theoretical treatment, and use this mode to mediate the MS interaction. For any practical operating conditions, the frequency difference between the COM mode and any of the other modes of an $N$ ion crystal is much larger than the Rabi frequency of the driving field \cite{James1997}, such that we can neglect the coupling to all other modes. For all experiments, the ions are initially Doppler cooled on the $4^2S_{1/2} \xrightarrow{} 4^2P_{1/2}$ transition, followed by sideband cooling of the COM mode. 
State readout is performed by fluorescence detection with a photomultiplier tube~\cite{schindler2013quantum}.

\subsection{Hamiltonian of the MS gate}

The MS gate is a commonly used method to generate entanglement in trapped-ion systems by exploiting a common vibrational mode of the ions. The gate is based on the application of a force that is dependent on the internal state of the ions. This force induces a periodic movement of the motional state of the ions in phase space.
At the end of the gate the motion is returned to its original state, but with an accumulated relative phase in the internal states. This interaction can then be used to create entanglement between the qubits.
The gate is experimentally implemented by the application of a bichromatic laser field which introduces four transition paths for a resonant two-photon process between the internal states, each of them going through a virtually excited intermediate state close to a motional sideband, as schematically depicted in Fig.~\ref{fig:MS_levels}(a). This process can be used to perform correlated spin-flips mediated by the common vibrational modes of the ions in the trap. The MS gate has desirable properties for a trapped-ion entangling gate, such as the ability to create entanglement between more than two ions with a single operation, and being independent of the initial motional state to first order. This last property guarantees robust functioning of the gate even with imperfect ground state cooling~\cite{Sorensen1999, sorensen2000entanglement, Kirchmair2009}. 
In combination with single-qubit rotations, the MS gate allows for universal quantum computation \cite{nebendahl2009optimal}.

In the following, we review how this gate is implemented and the experimental control parameters that need to be calibrated in order to obtain a high-fidelity realisation of the gate operation.

The desired action of the MS gate is an entangling operation acting on $N$ ions of the form
\begin{equation}
\label{eq:MS_operator}
    \mathrm{MS}_\varphi(\theta)=\exp\left(-i\theta S^2_\varphi\right),
\end{equation}
where 
\begin{equation}
    S_\varphi=S_y\cos(\varphi)+S_x\sin(\varphi),
\end{equation}
is the total spin operator in the direction defined by the angle $\varphi$ and
\begin{equation}
    S_\alpha=\frac{1}{2}\sum_{j=1}^N\sigma_{\alpha,j}, \quad \alpha=x,y,z
\end{equation}
with $\sigma_{x,j}$, $\sigma_{y,j}$ and $\sigma_{z,j}$ being the Pauli operators acting on the qubit encoded by the internal state of the $j$\textsuperscript{th} ion. By choosing $\theta=\pi/2$ one can use this gate to map computational basis states of $N$ qubits to maximally entangled states.

In order to derive the unitary evolution introduced by the MS gate, let us consider a system of $N$ ions in a linear trap driven by a bichromatic laser of the two frequencies, $\omega_1$ and $\omega_2$. The Rabi frequency, $\Omega(t)$, is assumed to be equal for all ions and can be time-dependent for a general pulse-shape of the laser.

This system may be described by the Hamiltonian
\begin{equation}\label{eqn:TotalHamiltonian}
\begin{alignedat}{2}
    &H= H_0 + &&H_\mathrm{int},\\
    &H_0 = \sum_{j=1}^N&&\frac{\omega_{eg}(t)}{2}\sigma_{z,j} + \omega_{sb} (a^\dagger a + 1/2), \\
    &H_\mathrm{int} = \sum_{j=1}^N&&\,\frac{\Omega(t)}{2}\left(\sigma^+_j + \sigma^-_j\right)\\& &&\cdot\Big(e^{i(\vec{k}_1\vec{x}_j-\omega_1 t+\varphi)}+e^{i(\vec{k}_2\vec{x}_j-\omega_2 t+\varphi)}+ h.c.\Big),
\end{alignedat}
\end{equation}
where $\omega_{eg}(t)$ is the transition frequency between the internal states $\ket{e}$ and $\ket{g}$, equal to a bare transition frequency $\omega_{eg,0}$ plus an AC Stark shift, $\omega_{AC}(t)\propto \Omega(t)^2$, due to the interaction of the laser with off-resonant atomic levels \cite{haffner2003precision,Kirchmair2009}; $\omega_{sb}$ is the frequency of the COM mode, which defines the distance of the motional sidebands (Fig.~\ref{fig:MS_levels}(a)) from the carrier; $a^\dagger$ and $a$ are the ladder operators related to the COM mode; $\sigma_j^+$ and $\sigma_j^-$ are the ladder operators acting on the internal states of the $j$\textsuperscript{th} ion; $\varphi$ is the phase of the two laser tones, which we consider to be equal; $\vec{k}_1$ and $\vec{k}_2$ are the wavenumbers of each laser tone.

We can describe the two laser frequencies $\omega_1$ and $\omega_2$ in terms of their symmetric, $\omega_d$, and asymmetric, $\omega_{a}$, detunings from the atomic transition frequency $\omega_{eg,0}$ (see Fig.~\ref{fig:MS_levels}) as: $\omega_1=\omega_{eg,0}+\omega_{d}+\omega_{a}$ and $\omega_2=\omega_{eg,0}-\omega_{d}+\omega_{a}$. The asymmetric detuning as well as the AC Stark shift of the electronic levels create a detuning $\omega_{cl}(t) = \omega_{AC}(t)-\omega_a$ of the mean value of the bichromatic frequencies from the carrier (Fig.~\ref{fig:MS_levels}(b)), known as center line detuning. Having $\omega_{cl}(t) \neq 0$ breaks both the condition $\omega_1 + \omega_2 = 2\omega_{eg}(t)$ of the two-photon resonance and the symmetry between the four paths involved in the gate (see Fig.~\ref{fig:MS_levels}(b)). Therefore, calibrating $\omega_{cl}(t)$ correctly is essential, and will be a central part of this work. With these definitions of $\omega_1$ and $\omega_2$, the interaction Hamiltonian $H_\mathrm{int}$ in Eq.~\eqref{eqn:TotalHamiltonian} becomes
\begin{align}
    H_\mathrm{int} =& \sum_{j=1}^N\frac{\Omega(t)}{2}\Big(e^{i(\vec{k}_1\vec{x}_j-(\omega_{eg,0}+\omega_{d}+\omega_{a}) t+\varphi)}\\&+e^{i(\vec{k}_2\vec{x}_j-(\omega_{eg,0}-\omega_{d}+\omega_{a}) t+\varphi)}+ h.c.\Big)\left(\sigma^+_j + \sigma^-_j\right).\nonumber
\end{align}
We can write $\vec{k}_i \vec{x}=\eta_{i}(a^\dagger+a)$ with $i=1,2$ where $\eta_i$ is the Lamb-Dicke parameter, and since $\omega_{d}\ll \omega_{eg}$ we can assume $\eta_{1},\eta_{2} \approx \eta$. We can simplify this Hamiltonian by assuming that we are in the Lamb-Dicke regime, $\eta\ll 1$, transforming to the interaction picture, introducing the sideband detuning, $\delta = \omega_{sb}-\omega_d$, and defining 
\begin{equation}
    \Lambda(t)\equiv\int_0^t\omega_{cl}(t')dt',
\end{equation}
that describes an unwanted accumulated phase due to the existence of a center line detuning during the gate operation. As a result, we obtain the following Hamiltonian
\begin{align}
    \label{eq:interaction_H}
    \hat{H}=-\eta\Omega(t)&(a^\dagger e^{i \delta t}+ae^{-i \delta t})\\&\cdot\left[S_y\cos(\varphi+\Lambda(t))+S_x\sin(\varphi+\Lambda(t))\right]\nonumber,
\end{align}
where we have used the rotating wave approximation to keep only the terms rotating with $\delta$ and ignore the other fast-rotating terms that go with $\omega_{sb}+\omega_d$ or $\omega_{eg,0}$.

In the experiment, we can realize the desired entangling gate from the Hamiltonian in Eq.~\eqref{eq:interaction_H} by adjusting the gate time $t_g$, sideband detuning $\delta$, center line detuning $\omega_{cl}(t)$, and phase $\varphi$. Let us consider first the case $\omega_{cl}(t) = 0$. Here the Hamiltonian can be integrated analytically to obtain the corresponding evolution operator
\begin{equation}
    \label{eq:MS_evolution_operator}
    \hat{U}(t) = D\big[\gamma(t) S_\varphi \big]\exp\big[i \theta(t) S^2_\varphi\big],
\end{equation}
where 
\begin{gather}
    \gamma(t) = i\eta \int_0^t  \Omega(t') e^{i t' \delta} dt',\\
    \theta(t) = \eta^2 \Im \int_0^t dt'\int_0^{t'}dt''\Omega(t')\Omega(t'') e^{- i \delta (t''-t')},
\end{gather}
and $D(z)=\exp(z a^\dagger - z^* a)$ is the displacement operator. Equation~\eqref{eq:MS_evolution_operator} allows us to choose the parameters of the gate in such a way that the result is the maximally entangling MS gate $\mathrm{MS}_\varphi(\pi/2)$.

First, in order to obtain the required entanglement between the qubits, we need to ensure that the rotation angle $\theta(t_g)$ at the end of the application of the gate satisfies $\theta(t_g)=\pi/2$. In the case of a constant pulse-shape this condition implies that 
\begin{equation}
    (\eta\, \Omega)^2\, t_g/\delta=\pi/2.
    \label{eq:ms_gate_condition}
\end{equation}

\begin{figure*}[t]
    \centering
    \includegraphics{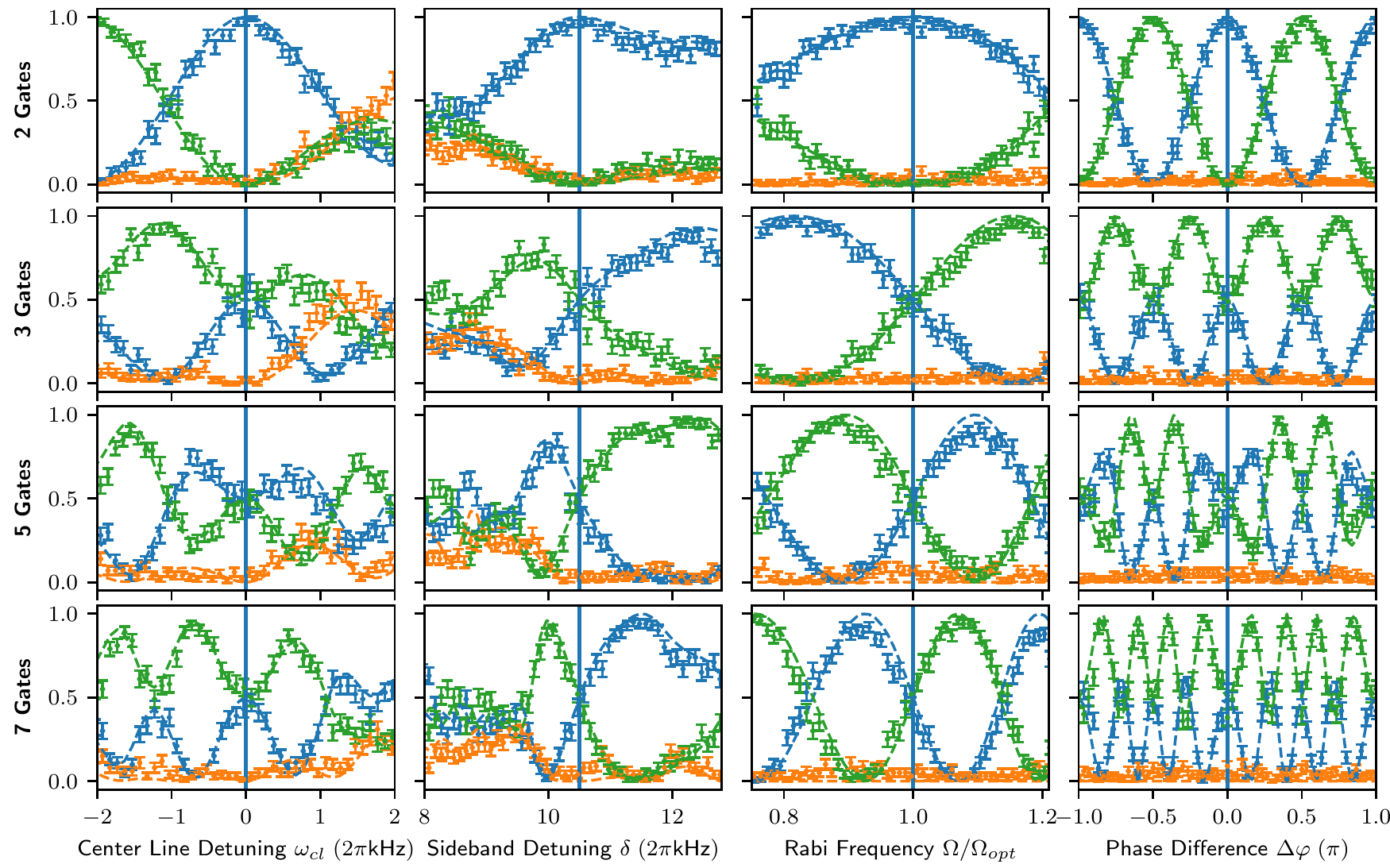}
    \caption{Populations of $\ket{g,g}$ (green), of $\ket{e,e}$ (blue), of $\ket{e,g}$ and $\ket{g,e}$ (orange) after the application of 2, 3, 5, and 7 gates with varying control parameters. For each scan the other parameters have been held constant at the optimal value for the MS gate. The error bars are estimated from shot noise. Dashed lines indicate the numerical simulations. The vertical lines mark the target value for each parameter.}
    \label{fig:parameter_scans}
\end{figure*}

Then, in order to avoid any residual entanglement between the internal states of the ions and their motional state, the sideband detuning needs to be chosen such that the quantity $\gamma(t)$ satisfies $\gamma(t_g) = 0$. This corresponds to closing a loop in phase space (as shown in the example in Fig.~\ref{fig:MS_levels}(a)), ensuring that the motional state returns to the initial state at the end of the gate. In the case of a constant pulse-shape the  condition $\gamma(t_g) = 0$ implies that 
\begin{equation}
    \abs{\delta} = 2\pi K /t_g
    \label{eq:ms_gate_condition_detuning}
\end{equation}
for any integer $K$ representing the number of loops in phase space the gate introduces (see Fig.~\ref{fig:MS_levels}(a)). In the following we will consider a single-loop gate, i.e., $K=1$.

In the previous discussion, we have considered the laser frequencies to be centered around the carrier, that is, $\omega_{cl}(t)=0$ at all times. However, since the AC Stark shift is time-dependent when we consider a time-dependent laser pulse, this cannot be satisfied at all times with constant laser frequencies. Therefore, in the following we will consider a time-independent center line detuning, $\omega_{cl}(t)=\omega_{cl}$, where we will aim to set it to zero. The error appearing from this approximation will be given by a non-zero value of $\Lambda(t_g)$. This has the effect of accumulating a phase, $\Delta \varphi=\Lambda(t_g)$, for each application of each consecutive gate, as can be seen from Eq.~\eqref{eq:interaction_H}. This cumulative phase shift can be detected by applying the MS gate more than once, and corrected by introducing a phase between consecutive gates.

\subsection{Experimental control parameters}
\label{sec:ModelMSParams}

In order to calibrate the MS gate we need to set the gate parameters 
\begin{equation}
\label{eq:exp_parameters}
    \boldsymbol{\Theta}=(\Omega, \,\omega_{cl}, \,\delta, \,\Delta\varphi),
\end{equation}
where $\Omega$ is the Rabi frequency, $\omega_{cl}$ the center line detuning,  $\delta$ the sideband detuning, and $\Delta\varphi$ the phase difference  of the gate as close as possible to their optimal values
\begin{equation}
    \boldsymbol{\Theta}_{opt}=\left(\Omega_{opt}=\frac{\pi}{\eta t_g}, \,\omega_{cl,opt}=0, \,\delta_{opt}=\frac{2\pi}{t_g}, \,\Delta\varphi_{opt}=0\right)
\end{equation}
as stated in the previous section, where the condition for $\Omega_{opt}$ is obtained from Eq.~\eqref{eq:ms_gate_condition} and Eq.~\eqref{eq:ms_gate_condition_detuning}. Since we do not have direct access to these parameters,  we rely on changing the control parameters of the gate $\boldsymbol{\Theta}_c = (t_g, \,f_{cl}, \,f_{sb}, \,\phi)$, i.e. the gate time $t_g$, the common frequency $f_{cl}= (f_r+f_b)/2$ is the mean frequency between the red ($f_r$) and blue ($f_b$) tone of the laser, the difference frequency $2f_{sb}=f_b-f_r$ of the bichromatic laser field, and the difference in the common phases between consecutive gates, $\phi$, of the two laser tones. 

These control parameters $\boldsymbol{\Theta}_c$ can be used to calibrate the gate in the following way:
By performing measurements of the populations of the ions after the application of a gate sequence we can obtain estimates that we denote as 
\begin{equation}
    \overline{\boldsymbol{\Theta}}=(\overline{\Omega}, \,\overline{\omega}_{cl}, \,\overline{\delta},
    \,\overline{\Delta \varphi}),
    \label{eq:parameter_estimates}
\end{equation}
of the current parameters $\boldsymbol{\Theta}$. The control parameters $\boldsymbol{\Theta}_c$ can then be adjusted to set the parameters $\boldsymbol{\Theta}$ closer to $\boldsymbol{\Theta}_{opt}$ given the estimates of the parameters: The time of the gate will be corrected as
\begin{equation}
    t_g \rightarrow t_g\frac{\Omega_{opt}}{\overline{\Omega}},\\
    \label{eq:update_power}
\end{equation}
where we choose to change the time of the gate instead of the laser power since it is easier to control in the experiment, while producing an equivalent correction.

The other corrections are implemented by subtracting the difference between the estimated and the ideal parameter value from the control parameter. In the case of the sideband detuning this is given by
\begin{equation}
    f_{sb}\rightarrow f_{sb} - \bar{\delta} + \delta_{opt},
    \label{eq:update_sideband}
\end{equation}
where $f_{sb}$ is the corresponding control parameter. 
As for the correction of the center line detuning, this is described by
\begin{equation}
    f_{cl} \rightarrow f_{cl}-\bar{\omega}_{cl} +\omega_{cl, opt}.
    \label{eq:update_centerline}
\end{equation}
Finally, the phase between consecutive gates is changed by
\begin{equation}
    \phi \rightarrow \phi - \overline{\Delta \varphi}+ \Delta \varphi_{opt}.
    \label{eq:update_phase}
\end{equation}
With this set of rules, we update the parameters of the MS gate to iteratively bring the parameters $\boldsymbol{\Theta}$ closer to $\boldsymbol{\Theta}_{opt}$.

\subsection{Validating the MS model}
\label{sec:model_comparison}

Next, we aim to verify that the theoretical description of the gate action agrees with the experiment. Since we are focusing on the effects of systematic parameter miscalibration, we are neglecting other error sources such as finite motional and spin coherence times, laser amplitude noise, resonant carrier excitation or unequal coupling strengths to the ions. As we target two qubit gates, all the following results presented are measured or calculated using a two ion crystal. 

In order to validate the Hamiltonian in Eq.~\eqref{eq:interaction_H} as a model we measure the outcome probabilities of a gate sequence while individually varying each control parameter. 
The remaining parameters are kept constant at their optimal values determined by manual calibration of the gate (Fig.~\ref{fig:parameter_scans}). 

The laser pulse is switched on adiabatically using a Blackman like shape~\cite{Schindler2008}, with a $4\mathrm{\mu s}$ shape time at the beginning and the end of the pulse to slowly increase the laser power to its maximum. The shaping ensures adiabaticity during the switch-on of the laser, preventing unwanted excitation of the carrier transition.
The gate time $t_g$ is defined as the duration of the laser pulse of the full width at half maximum.

We compare the measurements to the expectation values obtained from numerically integrating the Hamiltonian using the QuTiP software package \cite{Johansson2013}. 
Comparing the measurement results to numerical simulations with the same parameters, we calculate reduced $\chi^2$ values of $< 2.5$ for varying the center line detuning, sideband detuning and phase, while the Rabi frequency scans for $N_g=5$ and $N_g=7$ gates have higher $\chi^2$ values of up to 4. 
We attribute this to the AC Stark shift not being re-compensated for these measurements as the actual laser power was varied instead of adjusting detuning and gate time to control the Rabi frequency $\Omega$. 
While the $\chi^2$ are larger than one, these results do indicate that the Hamiltonian Eq.~(\ref{eq:interaction_H}) captures the effects of the parameter miscalibrations and is not dominated by unmodelled error sources. This suggests that the Hamiltonian is a viable model to describe our experimental system, and the results from numerically integrating the dynamics can be used as a probability landscape to perform Bayesian inference on. 

\section{Bayesian Inference}
\label{sec:BayesInf}

\begin{figure*}[t]
    \centering
    \frame{\includegraphics[width=\textwidth]{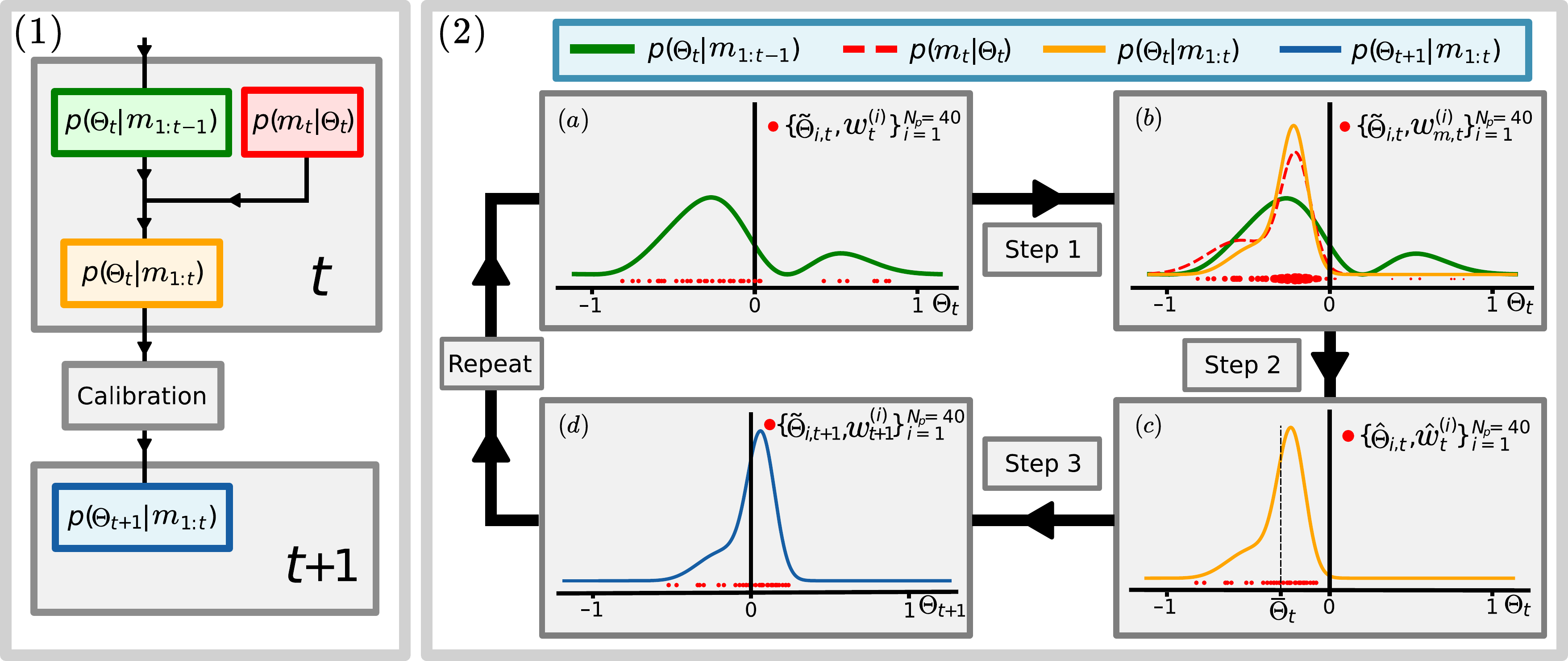}}
    \caption{(1) Representation of a cycle of the calibration process. The initial probability distribution at iteration $t$ with $t-1$ measurements, $p(\boldsymbol{\Theta}_t|m_{1:t-1})$, is updated by performing a new set of measurements, $m_t$, to obtain $p(\boldsymbol{\Theta}_t|m_{1:t})$. The estimates obtained from this probability distribution can then be used to calibrate the parameters, obtaining the initial probability distribution for iteration $t+1$. (2) Steps followed to perform an iteration of the particle filtering algorithm for a simplified case with only one parameter, $\Theta$. In (a) we have particles from the prior probability distribution $p(\Theta_t|m_{1:t-1})$ (green). A new set of measurements is performed to obtain the posterior probability distribution $p(\Theta_t|m_{1:t})$ (yellow) in (b) by using the likelihood of the outcome obtained for the measurement, $p(m_t|\Theta_t)$ (red). The weights are updated following Eq.~\eqref{eq:particle_update_rule}, where a bigger weight is visually represented by a bigger size of the particle. These weighted particles can be used to obtain the estimate of the parameter at this iteration, $\bar{\Theta}_t$, by using Eq.~\eqref{eq:mean_particles}. In (c), a resampling process is performed to obtain particles with equal weights that approximate particles from the posterior probability distribution. The previously obtained value $\bar{\Theta}_t$ is also shown here. Finally, in (d), $\bar{\Theta}_t$ has been used to calibrate the parameter $\Theta$, where for this example we are assuming that the ideal value of $\Theta$ is zero, therefore, the correction applied is to change the value of the parameter by $-\bar{\Theta}_t$. The probability distribution $p(\Theta_t+1|m_{1:t})$ (blue) consequently has an estimate of 0.}
    \label{fig:particle_filter_panel}
\end{figure*}

In order to estimate the values of the experimental control parameters that yield a high fidelity MS gate implementation, we need to determine their relation to the model parameters $\boldsymbol{\Theta}$. We are employing Bayesian inference for this task, since this framework allows us to straightforwardly incorporate information about our system obtained from measurement results using different gate sequences and control parameters $\boldsymbol{\Theta}_c$, experiment settings or previous calibration measurements. Additionally, this approach allows one to quantify the uncertainty on the estimated parameters, allowing us to monitor the progress of the calibration procedure and terminate the algorithm once the estimates reach sufficient accuracy.

Bayes' theorem \cite{lee2012bayesian} prescribes how to estimate the probability distribution of a set of parameters $\boldsymbol{\Theta}$, given a prior distribution before the measurement  $P(\boldsymbol{\Theta})$, a measurement with outcome $m$, and the likelihood of obtaining this measurement outcome, $P(m|\boldsymbol{\Theta})$, given a model of the process. 
The result is an updated posterior probability distribution 
\begin{equation}
\label{eq:bayes}
P(\boldsymbol{\Theta}|m)\propto P(m|\boldsymbol{\Theta})P(\boldsymbol{\Theta})
\end{equation}
up to a normalisation factor \cite{Ho1964ABA}. If we perform a new measurement we can apply Bayes' theorem again using the obtained posterior as a prior for the next measurement. This defines an iterative process (see Fig.~\ref{fig:particle_filter_panel}) in which the probability distribution after $T$ measurements is given by
\begin{equation}
    p(\boldsymbol{\Theta}|m_1,...,m_T)\equiv p(\boldsymbol{\Theta}|m_{1:T})\propto P(\boldsymbol{\Theta})\prod_{t=1}^T P(m_t|\boldsymbol{\Theta}),
    \label{eq:iterative_posterior}
\end{equation}
where we assume that all the measurement outcomes, $m_{t}$ for $t=1,...,T$, are independent from each other. For an increasing number of measurements we expect the degree of uncertainty of the parameters $\boldsymbol{\Theta}$ to decrease, and the process can be terminated if a desired limit of uncertainty for the estimates of $\boldsymbol{\Theta}$ is reached. 

\begin{figure*}[t]
\centering
\includegraphics{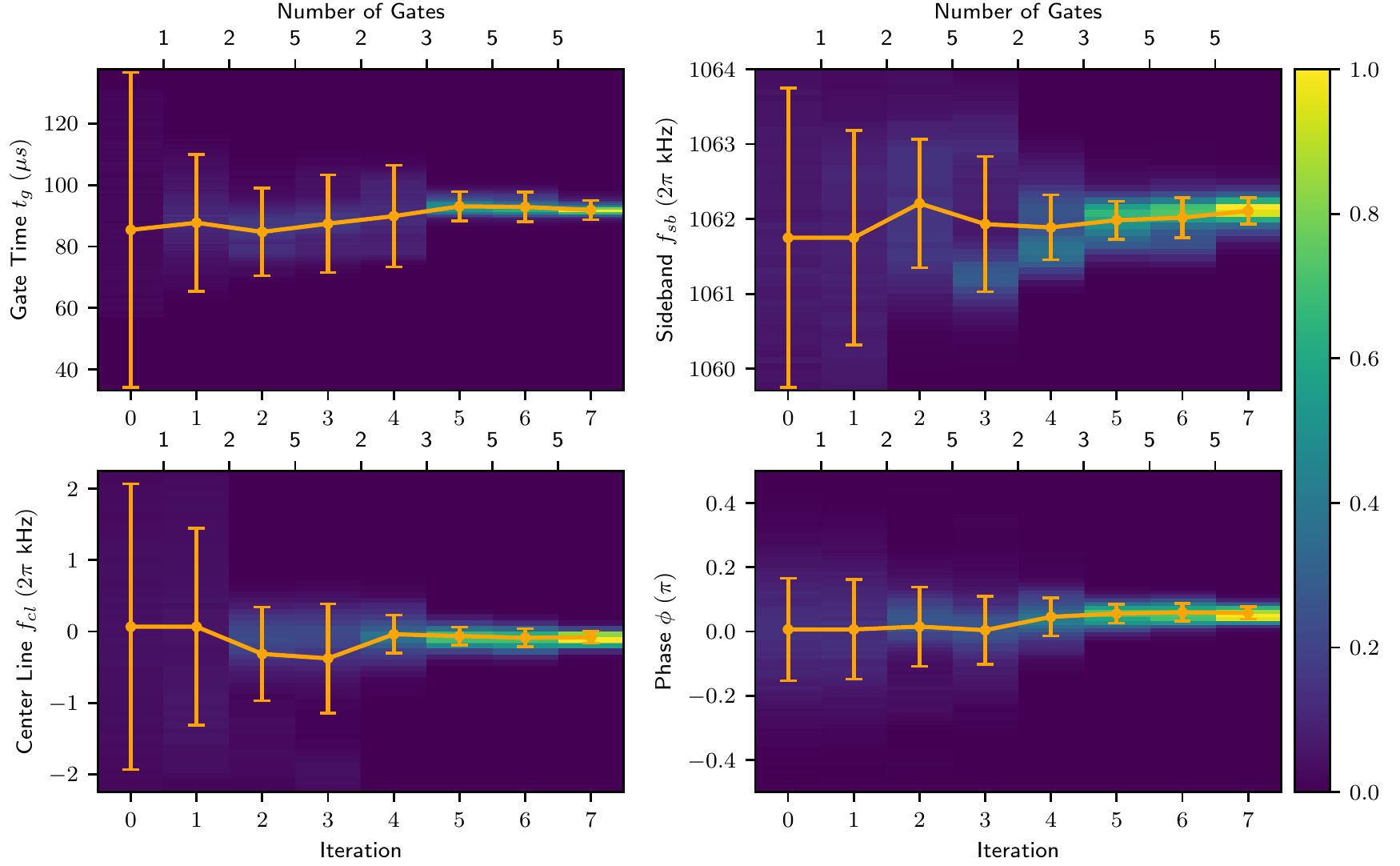}
\caption{Projections of the particle filter for the control parameters $\boldsymbol{\Theta}_c$ at each iteration of a single calibration run. The mean and deviation of the probability distribution of the control parameters, calculated using the relations in Eq.~\eqref{eq:update_power},\eqref{eq:update_centerline},\eqref{eq:update_sideband},\eqref{eq:update_phase}, is shown in orange. The probability density of each plot is normalized to its maximum. Each iteration uses the measurement outcomes of 100 repetitions of the experiment to update the probability density. As an initial prior a Gaussian  with widths of $\sigma_{\Omega}=0.2\cdot\Omega_{opt}$ Rabi frequency uncertainty, $\sigma_{\omega_{cl}}=2\cdot 2\pi \mathrm{kHz}$ center line uncertainty, $\sigma_\delta=2\cdot 2\pi \mathrm{kHz}$ sideband uncertainty and $\sigma_{\Delta\varphi}=0.16\pi$ phase uncertainty was chosen, which corresponds to typical experimental uncertainties after preliminary calibration. 
The marks at the top indicate the length of the gate sequence used for the measurements, chosen according to the variance minimization strategy described in Sec.~\ref{sec:measure_strategies}.} 
\label{fig:single_run_particle_filter_projections}
\end{figure*}

In order to calculate the posterior probability distribution of the parameters of the M{\o}lmer-S{\o}rensen gate using Eq.~\eqref{eq:bayes}, we need to know the probability of obtaining any measurement outcome $m$ for any given set of parameters $\boldsymbol{\Theta}$. 
These probabilities $P(m|\boldsymbol{\Theta})$ can be obtained by integrating the Hamiltonian in Eq.~\eqref{eq:interaction_H}. However, for $\omega_{cl}\neq 0$, no analytical solution is known that results in a closed-form expression for these probabilities, thus requiring to numerically study the effects of this type of miscalibration~\cite{Kirchmair2009}. 
As a consequence, we investigate a numerical treatment of the probability distributions.
In the following subsection we describe how the continuous probability distributions can be represented by discretized distributions instead. Afterwards, we will explain how these discretized distributions can be used to implement a Bayesian inference. A visual representation of our protocol is given in Fig.~\ref{fig:particle_filter_panel}.

\subsection{Representation of probability distributions}
We approximate the probability distributions using a so-called particle filter\footnote{The nomenclature of a filter stems from filtering estimates out of a stream of noisy measurement data} \cite{Mayne1966, Handschin1969, liu1998sequential}, where we replace the continuous probability function by a sum of $N_p$ weighted particles:
\begin{equation}
    p(\boldsymbol{\Theta})d\boldsymbol{\Theta} \approx\sum_{i=1}^{N_p} w^{(i)}\delta(\boldsymbol{\Theta} - \widetilde{\boldsymbol{\Theta}}_i) d\boldsymbol{\Theta},
    \label{eq:particle_filter}
\end{equation}
with $\delta(\boldsymbol{\Theta} - \widetilde{\boldsymbol{\Theta}}_i)$ being the Dirac delta function at the set of parameters $\widetilde{\boldsymbol{\Theta}}_i$ .
Each parameter set $\widetilde{\boldsymbol{\Theta}}_i$ is represented by a single particle with an associated weight $w^{(i)}$, and the set $\{\widetilde{\boldsymbol{\Theta}}_i, w^{(i)} \}_{i=1}^{N_p}$ forms the particle filter.
The weights $w^{(i)}$ satisfy $\sum_{i=1}^{N_p} w^{(i)}=1$, ensuring that the probability distribution is normalized. 

The Bayesian inference requires an initial prior, and we must choose an initial set of particles to represent this prior. We choose a known continuous probability density function as prior.
We can then initialize the particle filter from that probability function by randomly sampling $N_p$ times from the distribution and setting all the weights to $1/N_p$.

The density of this four-dimensional probability function describes the probability for a particular combination of Rabi frequency, $\Omega$, center line detuning, $\omega_{cl}$, sideband detuning, $\delta$, and phase difference, $\Delta\varphi$, to describe the parameters of the entangling MS gate.

\subsection{Bayesian Update} 
\label{sec:bayes_update}
The Bayesian update for the prior distribution, approximated by the particle filter $\{\widetilde{\boldsymbol{\Theta}}_i, w^{(i)}\}_{i=1}^{N_p}$, after performing a measurement (Step 1 in Fig.~\ref{fig:particle_filter_panel}) is implemented by updating the weights as given by
\begin{equation}
    w^{(i)}_{m} \propto P(m|\widetilde{\boldsymbol{\Theta}}_i) w^{(i)}.
    \label{eq:particle_update_rule}
\end{equation}
The weights $w^{(i)}_{m}$ are normalized such that $\sum^{N_p}_{i=1}w^{(i)}_{m} = 1$, ensuring that the filter represents a valid probability distribution.

The probabilities $P(m|\widetilde{\boldsymbol{\Theta}}_i)$ required to perform the update are obtained by numerical integration of the Hamiltonian in Eq.~(\ref{eq:interaction_H}) at each discrete sample point. Performing these numerical calculations in real time during an optimization run would, however, be prohibitively time consuming. The integration for a single point requires $\approx 1s$ on a CPU\footnote{CPU: Intel i5-4670S@3.10GHz}, which would result in a computation time of several hours to update 10000 particles which we use in our particle filter (See Sec.~\ref{sec:capture_range}). 

Instead, we precompute the outcome probabilities of a single experimental shot on an equally spaced 4 dimensional grid.
The outcomes between those grid points can then be calculated using an interpolation function. We use a spline interpolator \cite{walker2019quadcubic}, which allows us to better approximate the outcome probabilities without increasing the number of grid points required compared to using linear interpolation. 
For multiple gates the underlying grid uses 21 points for $\Omega,\, \omega_{cl},\, \delta$ and 25 for $\Delta\varphi$. The ranges of points are $\Omega= (1\pm 0.5)\cdot\Omega_{opt},\,  \omega_{cl}= 0\pm3.5\cdot  2\pi\mathrm{kHz},\, \delta= 10\pm5\cdot  2\pi\mathrm{kHz},  \,\Delta\varphi= 0\pm\pi$ for a gate duration of $100\mathrm{\mu s}$.
For the single gate grid we use a wider spacing with $\omega_{cl}= 0\pm7\cdot 2\pi\mathrm{kHz},\, \delta= 10\pm10\cdot 2\pi\mathrm{kHz}$.

The interpolation function can then be queried to receive the outcome probabilities at any set of parameters inside the region, with 10000 particles taking $\approx 100-500$ms to compute\footnote{CPU: Intel i7-6700K@4.00GHz}. This computation is now faster than the time required to acquire the experimental data from an iteration (See Sec.~\ref{sec:runtime}).
We additionally introduce a small amount $(1\%)$ of depolarizing noise into our model to account for experimental errors, in particular SPAM errors. This is necessary as noise is always present, preventing the expectation values $P(\ket{e,g})$ and $P(\ket{g,e})$ from going to 0, while the perfect model predicts arbitrarily small expectation values. The added depolarizing noise limits how much the likelihood can be adjusted by a measurement of $\ket{g,e}$ or $\ket{e,g}$.

\begin{figure*}[t]
    \centering

    \includegraphics{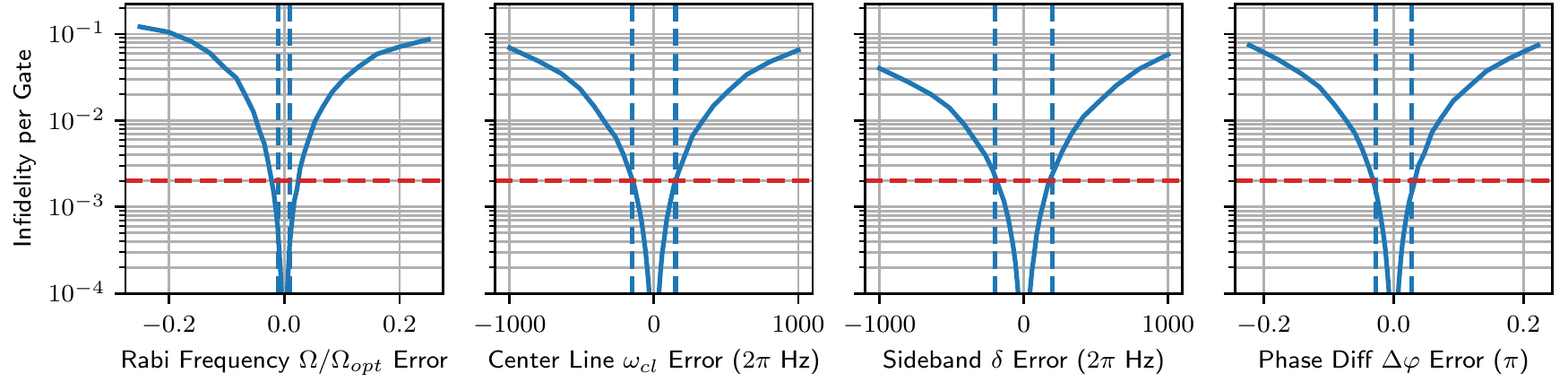}
    \caption{Expected Randomized Benchmarking infidelity as function of individual parameter miscalibrations for a $100\mu s$ gate. These curves inform the choice of target thresholds (blue dashed lines), an infidelity of $2\cdot 10^{-3}$ (red dashed line) for each parameter has been chosen as a threshold. The fixed parameters were kept at $\Omega=\Omega_{opt}$, $\omega_{cl}=0$, $\delta=10\cdot 2\pi\mathrm{kHz}$, $\Delta\phi=0$.
    }
    \label{fig:rb_infidelity}
\end{figure*}

\subsection{Parameter Estimation}
The particle filter allows us to straightforwardly extract statistical information from the distribution, most notably the mean and the variance of the probability distribution,
\begin{align}
    \label{eq:mean_particles}
    \overline{\boldsymbol{\Theta}} &= \sum_{i=1}^{N_p} w_m^{(i)} \widetilde{\boldsymbol{\Theta}}_i, \\
    \textrm{Var}(\boldsymbol{\Theta}) &= \sum_{i=1}^{N_p} w_m^{(i)} (\widetilde{\boldsymbol{\Theta}}_i-\overline{\boldsymbol{\Theta}})^2.
    \label{eq:particle_variance}
\end{align}
These estimates can be used during the algorithm to monitor progress and inform decisions about the next experimental setting (see section \ref{sec:measure_strategies}).

\subsection{Resampling}

We are expecting the variance of the probability distributions to decrease as information from more measurements is added, improving the estimates of the parameters.
This poses a known problem for the particle filter as we repeatedly update the weights, since it leads to many particle weights going to zero while simultaneously having only a few particles with high weights. 
This situation causes both unnecessary computations to update the weights on particles that represent a small probability and thus contribute little to the estimates, and under-sampling of the distribution around the particles with high probability, limiting the final precision of the estimate. We counteract this problem by a so-called resampling procedure \cite{gordon1993novel, Rubin1987}, where we generate a new set of particles with equal weights to represent the same probability distribution (Step 2 in Fig.~\ref{fig:particle_filter_panel}).
\begin{equation}
    \{\widetilde{\boldsymbol{\Theta}}_i,\, w_m^{(i)} \}_{i=1}^{N_p} \rightarrow \{\hat{\boldsymbol{\Theta}}_i,\, \hat{w}^{(i)}=1/N_p \}_{i=1}^{N_p}
\end{equation}
We use the Liu-West algorithm \cite{liu2001combined} for this task, which first generates a new set of particles by randomly sampling with replacement from the old weighted particles. 
This already generates a set of equally weighted particles, but particles cannot explore new locations since they are up to this point duplicates of particles of the old set. 
To lift this degeneracy every particle of the filter is moved towards the mean of the filter by a constant. A random perturbation is then applied to each particle, with the values of the perturbation sampled from a normal distribution. The covariance of this normal distribution is chosen such that the mean and covariance of the particle filter is preserved.

\subsection{Feedback}

The last step (Step 3 in Fig.~\ref{fig:particle_filter_panel}) remaining in our parameter estimation is to transform the posterior probability distribution into the prior distribution for the next iteration. We use the estimates from the particle filter (Eq.~\eqref{eq:particle_variance}) to calculate the required adjustments of our experimental control parameters for a perfect gate using Eqs.~\eqref{eq:update_power}-\eqref{eq:update_phase}, given the current knowledge of our parameter estimates (Eq.~\eqref{eq:parameter_estimates}). 

We also apply these corrections to the positions of our particles such that the expectation values of the new prior fulfill the relations for a perfect gate.
These new values can now be used in the next iteration of the algorithm, thus improving the estimates of the parameters iteratively. An example of this behaviour estimating the four MS gate parameters is shown in Fig.~\ref{fig:single_run_particle_filter_projections}, where the probability distributions are mapped to the experimental control parameters $t_g,\, f_{sb},\, f_{cl},\,$ and $\phi$. At each update, 100 experimental shots are performed using the estimates for the control parameters by the previous iteration and the particle filter is then updated according to the measurement results. Over several iterations the probability distribution narrows, reducing the uncertainty on the estimates (shown in orange), and the changes to the control parameters become smaller between iterations. 

\section{Calibration algorithm}
\label{sec:AutoAlgo}
We now proceed to use the Bayesian estimation framework introduced previously to calibrate all four key parameters that determine the performance of the M{\o}lmer-S{\o}rensen gate in our experiment.
We introduce a stopping criterion that relates the parameters to gate infidelities.
We describe the selection process of the measurement settings to improve convergence of the Bayesian parameter estimation protocol and investigate the effect on the experimental run time of the algorithm. 
The choice of measurement setting determines the amount of information that will be gained, and can have significant effects on the number of measurements required \cite{Huszar2012, Kravtsov2013, Wiebe2016, Granade2017, Martinez-Garcia2019}.
We finally experimentally verify the performance of the algorithm by checking the consistency of the final gate parameters returned by the algorithm.

\subsection{Stopping criterion for the algorithm}
\label{sec:thresholds}

\begin{figure}
    \centering
    \includegraphics{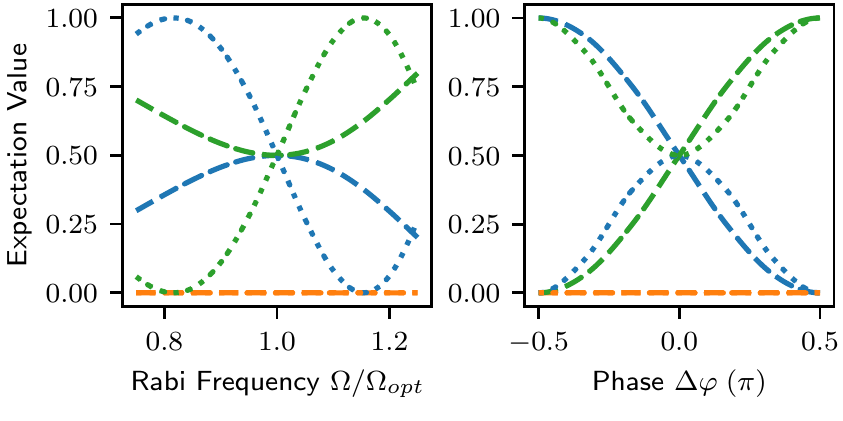}
    \caption{Expected outcome probabilities of $P(\ket{g,g})$ (green), $P(\ket{e,e})$ (blue) and $P(\ket{e,g}) + P(\ket{g,e})$ (orange) for a $3\; \mathrm{MS}_0(\frac{\pi}{2})$ gate sequence (dotted) and for a $\mathrm{MS}_0(\frac{\pi}{2})\mathrm{MS}_\frac{\pi}{4}(\frac{\pi}{2})$ sequence (dashed). The first gate sequence is first-order insensitive to phase miscalibrations as the local minimum of the outcome probabilities around zero leads to only small variations in the likelihood of possible outcomes, and its symmetry around zero does not allow to discriminate the sign of the miscalibration. 
    Similar arguments can be used to see that this sequence is first-order-sensitive to Rabi frequency miscalibrations. For the second gate sequence the relation is inverted, being sensitive to phase miscalibrations at the cost of first-order-insensitivity to the Rabi frequency. The qualitative behaviour of the sensitivity of the center line matches the sensitivity of the phase difference, while the behaviour of the sideband detuning matches the Rabi frequency.}
    \label{fig:setting_sensitivity}
\end{figure}

\begin{figure*}[t]
    \centering
    \includegraphics{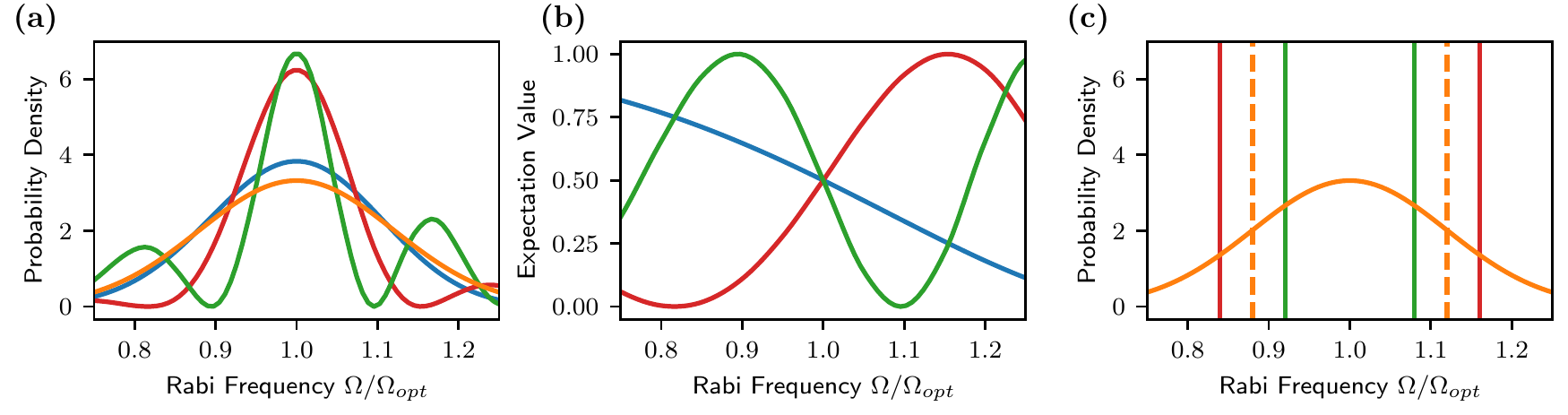}

    \caption{ Illustration of the decision-making process on which gate sequence to use in the next measurements. 
    The left plot (a) illustrates the variance minimization strategy for the prior given in orange and the expected posterior probability distributions calculated using the expected outcomes. 
    We compare the standard deviations (dashed lines) of the distributions to select the smallest for next measurement setting. In this example the distribution for 3 gates is the narrowest and is thus chosen. 
    The plot (b) shows the expected outcome probability $\textrm{Var}(\boldsymbol{\Theta}_s)$ for the $\ket{g,g}$ state according to the model for each setting under consideration ($N_g=1$ blue, $N_g=3$ red, $N_g=5$ green with $\Delta\varphi_{target}=0$) depending on the Rabi frequency $\Omega$. In this case the outcome probability for $P(\ket{e,e})=1-P(\ket{g,g})$ with population in the remaining states being zero. 
    In the right plot (c) the same prior is evaluated using the thresholded strategy. We only consider the standard deviation of the prior (dashed lines), with the width of the prior being larger than the 5 gate threshold (green) but below the 3 gate threshold (red), leading to the 3 gate setting being chosen. The single gate setting does not have any thresholds, being the fallback option if no other setting passes the thresholds.} 
    \label{fig:strategy_settings}
\end{figure*}

We want to determine the parameters of the gate with sufficient accuracy to perform operations with a target fidelity. The calibration procedure thus requires a stopping criterion to decide when the routine has reached the necessary accuracy. 
Entangling gates are usually not quantified by the uncertainty in their control parameters, but by their fidelity \cite{Jozsa1994} compared to the ideal expected output state. By averaging fidelities obtainable for various inputs states and unitary gate sequences, a more relevant quantity for performing an algorithm, the average gate error, can be extracted \cite{Gaebler2012,Emerson2005,Magesan2011}. 
The fidelity of any real gate is limited by noise processes. It is thus sufficient to determine the parameters to an accuracy so the present stochastic noise is the limiting factor of the gate fidelity. 
In order to relate miscalibrated parameters to gate fidelity, we simulate randomized cycle benchmarking~\cite{Erhard2019} with mis-set parameters. We consider the effect of the parameters individually, with the resulting infidelities shown in Fig.~\ref{fig:rb_infidelity}. For each parameter we can then define a threshold $T_\Theta$ which defines an acceptable region in which a miscalibration is no longer expected to significantly influence the gate performance. In our case we choose 
a threshold of $2\cdot 10^{-3}$ for the infidelity caused by a miscalibration of that single parameter, leading to a choice of thresholds on the individual parameters of $T_\Omega=0.02\cdot\Omega_{opt},\; T_{\omega_{cl}}=150\cdot2\pi \mathrm{Hz}, \; T_\delta=200\cdot2\pi \mathrm{Hz},\; T_{\Delta\varphi}=0.028\pi$.
We proceed to run the calibration algorithm until the particle filter converges in all four dimensions to an uncertainty below the thresholds. 
Assuming the parameters are normally distributed with the standard deviation of the distributions equal to the thresholds, we expect from simulation a median infidelity of $\approx 5\cdot 10^{-3}$, but we expect that correlations between the errors in the parameters can significantly affect the expected infidelity (see Sec.~\ref{sec:confirm}).

\subsection{Selection of measurement settings}
\label{sec:measure_strategies}
We have a choice of the sequence of gates for which we want to perform a measurement. Different gate sequences affect the control parameters differently. As an extreme example, a single gate has no dependency on the phase difference $\Delta\varphi$, a measurement of a single gate thus does not provide any information about that parameter.
We thus require a strategy for selecting suitable measurement settings that ideally maximize the amount of information gained from performing a certain measurement.

We restrict our considerations to sequences of MS gates without any local operations. 
Besides the number of gates $N_g$ we can also intentionally introduce a phase difference $\Delta\varphi_{target}$ between the gates.
We restrict the phase difference settings to either $\Delta\varphi_{target}=0$ or $\Delta\varphi_{target}=\pm \pi/4$, corresponding to either consecutive $\mathrm{MS}_0(\frac{\pi}{2})$ gates or a sequence of $\mathrm{MS}_0(\frac{\pi}{2})\,\mathrm{MS}_\frac{\pi}{4}(\frac{\pi}{2})\,...\,\mathrm{MS}_{(N_g-1)\frac{\pi}{4}}(\frac{\pi}{2})$ gates. We chose these two types of sequences as they change which parameters the sequence is most sensitive to (Fig.~\ref{fig:setting_sensitivity}).

Applying more gates increases the sensitivity of the sequence as the peaks of the likelihood function $P(m|\boldsymbol{\Theta})$ become narrower, resulting in an increased first-order sensitivity to miscalibrations. 
At the same time, the peaks of the likelihood function become more closely spaced, which for a wide prior can lead to a multi-modal posterior distribution which causes slow convergence. Additionally, the effect of unmodelled error sources such as decoherence increases with additional gates, limiting the total length of the sequence used in the optimisation.  
We thus need to choose the number of gates $N_g$ we apply, as well as the targeted phase difference between consecutive gates.

These gate sequences form a set of possible measurement settings. We work with two approaches for choosing from this set, which we present in the following: 

\begin{figure*}[t]
    \centering
    \includegraphics{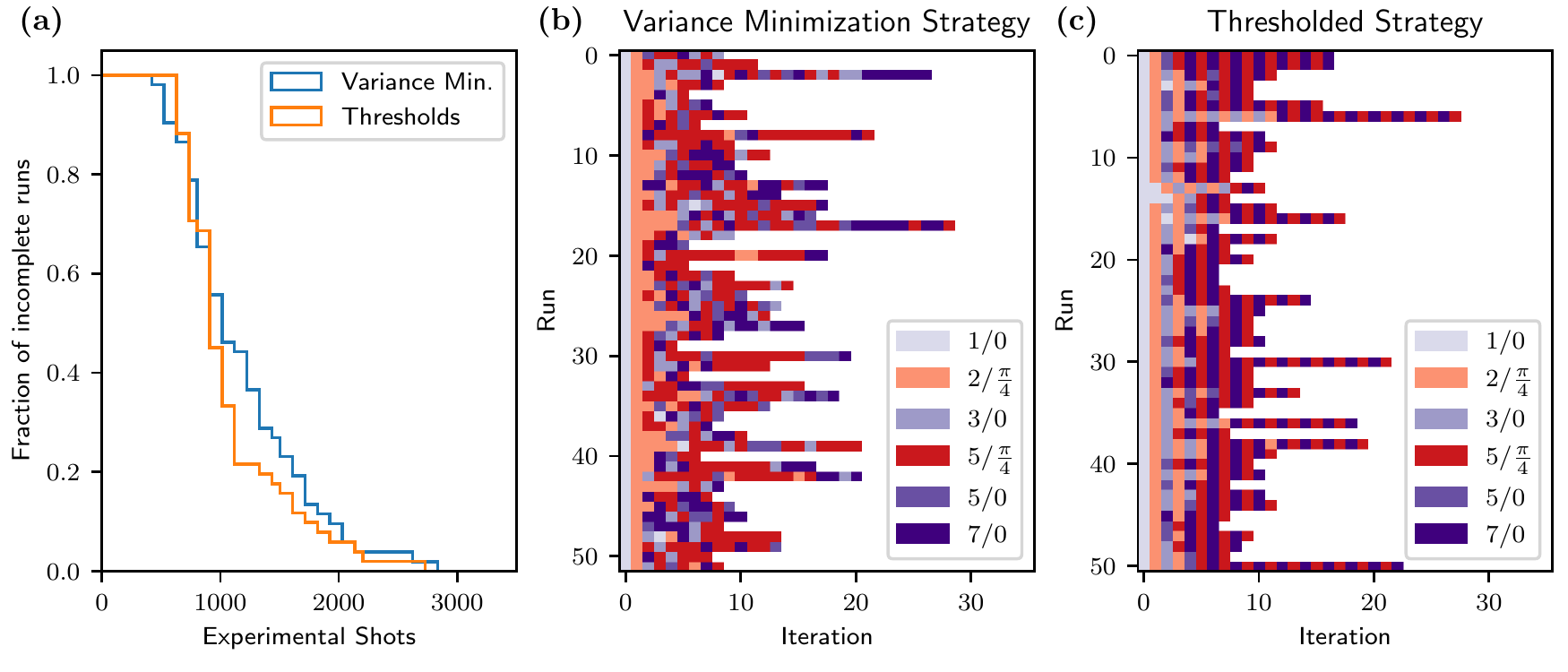}
    \caption{The number of experimental cycles required to reach the target thresholds is shown as a fraction of all runs in (a). The thresholded strategy takes $1100 \pm 500$ cycles on average compared to $1200 \pm 500$ for the variance minimization strategy. The figures (b) and (c) show the measurement settings used during the calibration runs for the respective strategies, with the y axis corresponding to independent runs and the x axis to iterations of the algorithm during a calibration run, at each iteration 100 measurements were performed with the chosen setting. 
    The colors indicate the measurement setting used at a given iteration of a run, characterized by the number of gates $N_g$ (first number in the legend) and the selected phase difference $\Delta\varphi_{target}$ (second number) used for the gate sequence.} 
    \label{fig:gates_used_over_run_time}

\end{figure*}

\subsubsection{Variance Minimization Strategy}
To maximize the information gained on our parameters from a measurement we aim at performing a measurement that decreases the variance of our posterior distribution as much as possible. 
For each measurement setting $s$ we can predict the posterior estimate $\overline{\boldsymbol{\Theta}}_{m_j, s}$ and variance $\textrm{Var}(\boldsymbol{\Theta}_{m_j, s})$ for each measurement outcome $m_j$ by first calculating the weights for each possible outcome of the measurement with Eq.~\eqref{eq:particle_update_rule} using the corresponding probabilities $P_s(m, \boldsymbol{\Theta})$
\begin{align}
    \overline{\boldsymbol{\Theta}}_{m_j, s} &= \sum_{i=1}^{N_p}  P_s(m_j|\boldsymbol{\Theta}_i) w_{\boldsymbol{\Theta}_i}^{(i)},\\
    \textrm{Var}(\boldsymbol{\Theta}_{m_j, s}) &= \sum_{i=1}^{N_p}  (\boldsymbol{\Theta}_i - \overline{\boldsymbol{\Theta}}_{m_j, s} )^2 P_s(m_j|\boldsymbol{\Theta}_i) w_{\boldsymbol{\Theta}_i}^{(i)}.
\end{align}

We can estimate the total probability of an outcome by averaging over the particle filter 
\begin{equation}
    \langle m_{j, s} \rangle = \sum_{i=1}^{N_p} P_s(m_j|\boldsymbol{\Theta}_i) w_{\boldsymbol{\Theta}_i}^{(i)}.
\end{equation}
By weighing the variance with the expected probability for that outcome we can calculate the most likely variance $\textrm{Var}(\boldsymbol{\Theta}_s)$ for a given measurement setting given the current knowledge before actually measuring, 
\begin{equation}
   \textrm{Var}(\boldsymbol{\Theta}_s) =\sum_{j} \textrm{Var}(\boldsymbol{\Theta}_{m_j, s}) \langle m_{j, s} \rangle 
\end{equation}
We aim to minimize our variance to improve the estimate of all our parameters, which is not straightforward for four separate variances for the parameters, as different measurement settings do not affect all parameters equally. We thus employ a heuristic to calculate a score for each measurement setting for which we normalize each variance with its target threshold $T_\Theta$ defined in Sec.~\ref{sec:thresholds}  and sum up all the normalized variances to get a score $X_s$ for the measurement setting 
\begin{equation}
    X_s = \sum_\Theta \frac{\textrm{Var}(\Theta_{s})}{T_\Theta^2}.
\end{equation}
We then apply the measurement setting with the lowest score to the experiment. The process is illustrated for 1 dimension in Fig.~\ref{fig:strategy_settings}. 

We need to calculate the outcome probabilities at each particle location for each measurement setting, which is computationally expensive. We interpolate probabilities from precomputed simulations to speedup this calculation, but the interpolation remains the most time intensive operation in our classical computations. 
While calculating the probabilities is an unavoidable step for performing the Bayesian update of the particle filter as discussed in Sec.~\ref{sec:bayes_update}, predicting the variances requires running this calculation for each measurement setting under consideration. This significantly increases the total classical computational overhead. 

\subsubsection{Thresholded Strategy}

In order to reduce the computational cost of finding the next measurement setting we present a simple heuristic as an alternative to the variance minimization strategy described above. For this we want to look at which settings are considered optimal by the Variance Minimization strategy. In practice this means selecting the setting that uses the most gates while avoiding a multi-modal posterior distribution. 
These multi-modal posteriors are caused by the likelihood function having two maxima over the range of the prior. To avoid them, ideally the chosen measurement settings would yield outcome probabilities that have only one likely parameter value. While it is impossible to simultaneously fulfill this for all parameters, we can still attempt to limit the amount of possible maxima of the likelihood function with the right choice of measurement setting.
To accomplish this, we want to impose a maximal requirement on the width of the prior before using a specific measurement settings.

We examine the expected outcomes of the model restricted to one dimension for each parameter and for each experimental setting, while the other parameters are set to their optimal values. 
We then choose the local extremum closest to the optimal value in the outcome probabilities, and define a threshold as the distance of the extremum to the center for each measurement setting. 
While this ignores possible correlations of the probability distribution between the parameters, we have found in practice that these thresholds are already sufficient to avoid multi-modal posteriors. 

We split the settings into first-order Rabi frequency $\Omega$ sensitive settings and into first-order phase difference $\Delta\varphi$ sensitive settings depending on whether the $\ket{e,e}$ and $\ket{g,g}$ populations have a local extremum exactly at the optimal parameter value or a linear slope in either phase or Rabi frequency respectively (see Fig.~\ref{fig:setting_sensitivity}). 

These thresholds are now used to define the measurement strategy that will be used by the calibration algorithm: 
On each iteration we alternate between selecting from either Rabi frequency or phase first-order-sensitive settings, comparing the defined thresholds to the current variances to ensure that the variance of both Rabi frequency $\Omega$ and phase difference $\Delta\varphi$ are being improved.
We now compare the variance of our prior with the thresholds, and discard any setting for which the variance exceeds a threshold in any parameter. Among the remaining settings we pick the setting with the highest number of MS gates (For an example see Fig.~\ref{fig:strategy_settings}). 
This gives us a way to select measurement settings while avoiding the computational overhead incurred by the Variance Minimization strategy, with the time needed to evaluate the strategy ($\ll 1$ms) being negligible compared to the time required to update the particle filter (Sec.~\ref{sec:bayes_update}). 

\subsection{Run time of algorithm}
\label{sec:runtime}

A single experimental shot only takes $\approx 10$ ms, while computing the particle filter update requires $\approx 100-500$ms.
To balance the classical computational effort for updating the particle filter and the latency in updating the control system with the duration of the experiment, we run 100 experimental cycles with the same measurement setting before updating the particle filter and choosing the next experimental setting.
The variance minimization strategy additionally needs time ($\approx 0.6-3$ s) for selection of the next measurement setting, while the time required to evaluate the thresholded strategy is negligible.

In Fig.~\ref{fig:gates_used_over_run_time} (a) we show the number of individual experiments required to reach the target thresholds, giving a comparable number of experimental shots for the variance minimization strategy ($1200\pm500$ shots) and the thresholded strategy ($1100\pm500$ shots). 
Fig.~\ref{fig:gates_used_over_run_time} (b) and (c) display the measurement settings used during these calibration runs. The variance minimization strategy focuses on either the phase or the Rabi frequency sensitive setting, trying to reduce the largest variance component before switching. The thresholded strategy instead alternates between the settings of different sensitivity. Both settings increase the number of gates on later iterations as the variances decrease. While the order of which measurement setting is chosen is different between the strategies, the number of experimental shots required is comparable.

From this we can conclude that the thresholded strategy is a good heuristic to select measurement settings compared to the variance minimization strategy.

Taking the computational overhead into account, the thresholded strategy has the advantage, completing with average run time of $41\pm 17$ s compared to an average of $60\pm 26$ s for the variance optimization. Averaged over both strategies we find that we require $1200\pm500$ experimental shots to reach our target thresholds. 
A strict quantitative comparison to all other possible optimisation strategies is difficult to define. Using our Bayesian algorithm we achieve a greater than 30\% speedup compared to manual iteration over 1D parameter scans as defined in reference \cite{Akerman2015} even whilst assigning optimistic run time assumptions to these scans. This comparison is fully discussed in Appendix
\ref{appendix:manual_tuneup}.

\subsection{Capture Range}
\label{sec:capture_range}

\begin{figure}[tb]
    \centering
    \includegraphics{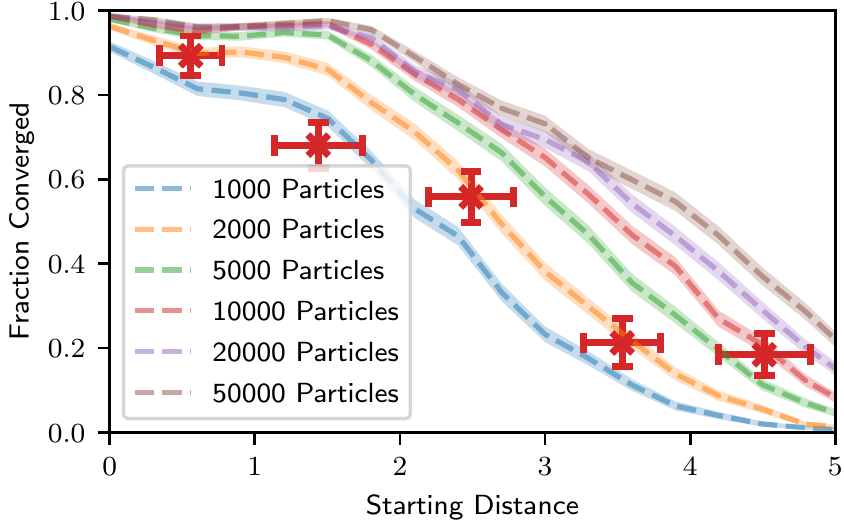}
    \caption{ Capture Range of the algorithm. The width of the initial Gaussian prior has been held constant at $\sigma_{\Omega}=0.2\cdot\Omega_{opt}$ Rabi frequency uncertainty, $\sigma_{\delta}=1\cdot2\pi\mathrm{kHz}$ sideband uncertainty, $\sigma_{\omega_{cl}}=1\cdot2\pi\mathrm{kHz}$ center line uncertainty and $\sigma_{\Delta\varphi}=0.33\pi$ phase uncertainty. 
    The starting distance has been calculated using Eq.~\eqref{eq:starting_dist}. The fraction captured represents the number of runs that terminate within two times the target thresholds of the optimal parameter values. For each distance, 2000 tuneups were run.
    For 10000 particles we additionally measured the capture range experimentally, where the results are represented with red squares. The error bars correspond to the statistical deviation of the data points.
    }
    \label{fig:capture_range}
\end{figure}

\begin{figure*}[t]
    \centering
    \includegraphics{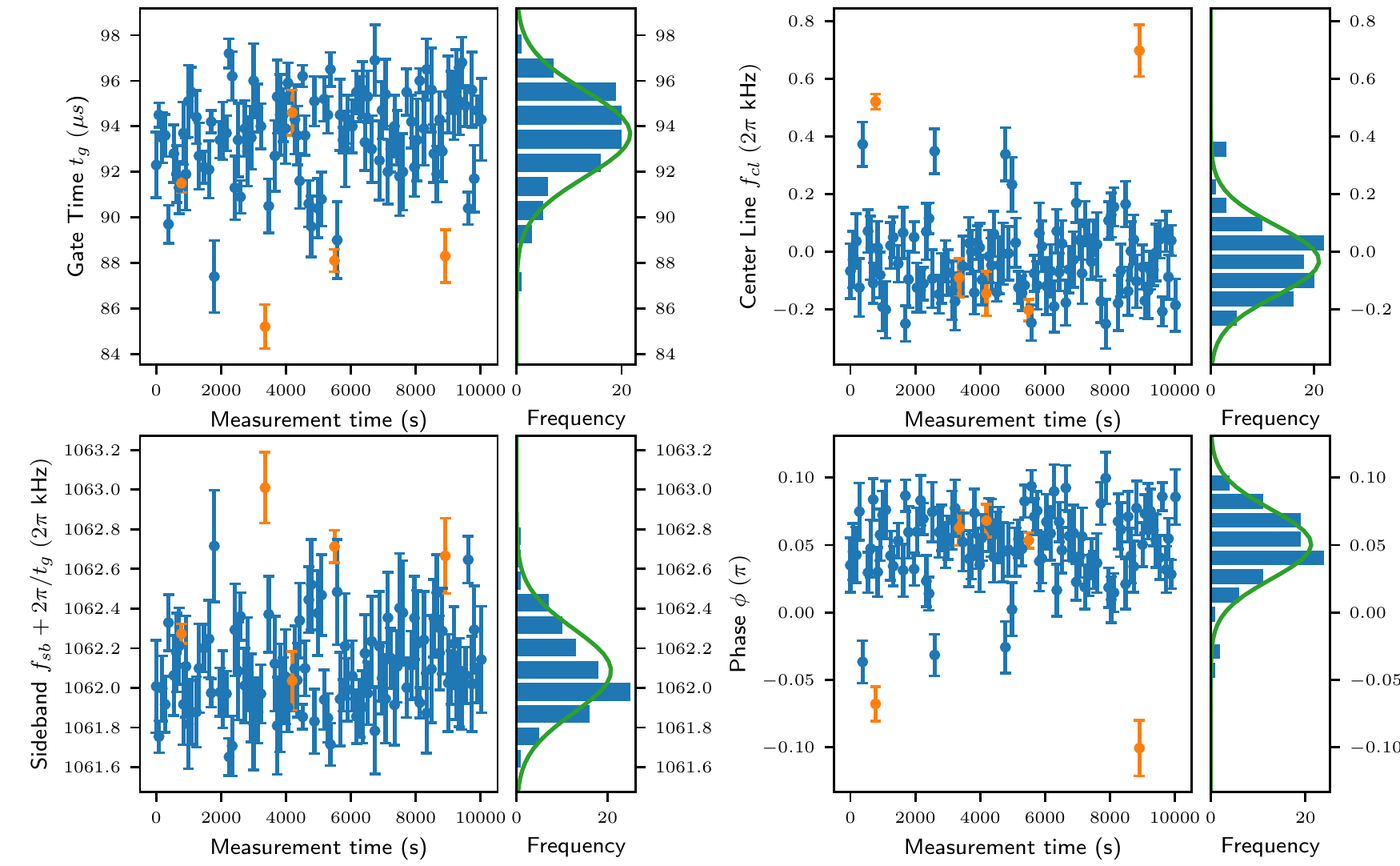}
    \caption{Final parameter estimates and uncertainties for 103 iterations of tuning up the M{\o}lmer-S{\o}rensen gate versus time elapsed. The orange points are runs rejected by the confirmation criterion described in Sec.~\ref{sec:confirm}, corresponding to 5(2)\% of all runs. Removing the rejected points reduces the standard deviation of the distributions of final parameter estimates between 13-20\%. Green curves show a Gaussian distribution fitted to a histogram the distribution of final parameter values.
    }
    \label{fig:endpoints_consistency}
\end{figure*}

Any Bayesian procedure requires a choice of an initial prior, where in our case we consider normal distributions. It is reasonable that if the true values are unlikely given our initial prior, that is, if the true values are too far away from the center of our initial prior, then the routine will fail to find them. In this section we will study the dependence of the failure rate of our algorithm on this initial distance in parameter space. We will also study the dependence of the success rate of the algorithm on the number of particles in the particle filter, since a larger number of particles allows for a more accurate sampling and, therefore, a better approximation of the probability distributions under study. This increase in the number of particles comes with the cost of a linear growth in the classical computing time required by the algorithm. 

We study these effects by first obtaining a normalized distance, $D_{start}$, of the initial values of the parameters from their target values using the widths of the prior
\begin{equation}
    D_{start}^2 = \sum_\Theta         \frac{(\Theta_{start}-\Theta_{opt})^2}{\sigma_\Theta^2}
    \label{eq:starting_dist}
\end{equation}
with $\sigma^2_\Theta$ being the initial variance of the prior projected onto parameter $\Theta$, with $\Theta_{start}$ the starting value and $\Theta_{opt}$ the optimal value for that parameter.

We generate random combinations of initial gate time, center line frequency and sideband frequency with the chosen starting distance and then run simulated calibration runs. The phase was not varied as the space is periodic, and thus limits how far the phase values can be from the optimal value. 
The fraction of calibration runs that converged is shown in Fig.~\ref{fig:capture_range}, where we consider any run that ends up closer than twice the termination thresholds of the stopping criterion (Sec.~\ref{sec:thresholds}) in all parameters a success. 
Since for each parameter the estimated error is below the threshold, for a Gaussian distribution at least 95\% of results should be between these thresholds. Due to the statistical nature we do not expect a perfect success rate even if the particle filter were to perfectly approximate the continuous probability function. 

We find that for 2000 particles and less in the particle filter the capture range is reduced as compared to the higher particle numbers, and performance is decreased even at starting distance of 1 compared to the higher particle numbers. 
This is expected as a particle filter with a too low number of particles will lead to an undersampling of the probability distribution, which in turn results in a poor approximation of this probability distribution. As a consequence discretisation artifacts will appear, such as the particle filtering not being able to track well enough the peaks of the distribution.

We choose 10000 particles as our default value to work with since in simulation the gains for using more than 5000 particles decrease significantly. 

Experimentally, we find a reduction of 10-25\% of the success rate compared to the simulations using the same sample size for starting distances $<2$ due to effects not included in the model affecting the algorithm. For larger distances, the success rate is comparable between experiment and simulation.

While we are unable to pin down the exact cause of failures, we hypothesize that these events may be due to transients in the control fields. 

These effects may lead to a difference between our estimated likelihoods and the true ones, which then suppress the posterior probability distribution around the true values. This can then shift the parameter estimates away from the true values, but also lead to underestimating the variances compared to the true error. 
The algorithm might then use unsuitable measurement settings, or prematurely terminate if the variances are small enough to fulfill the stopping criterion. These effects can then lead to a final parameter estimates of the algorithm far from the ideal values.

\begin{figure}[hbt]
    \centering
    \includegraphics{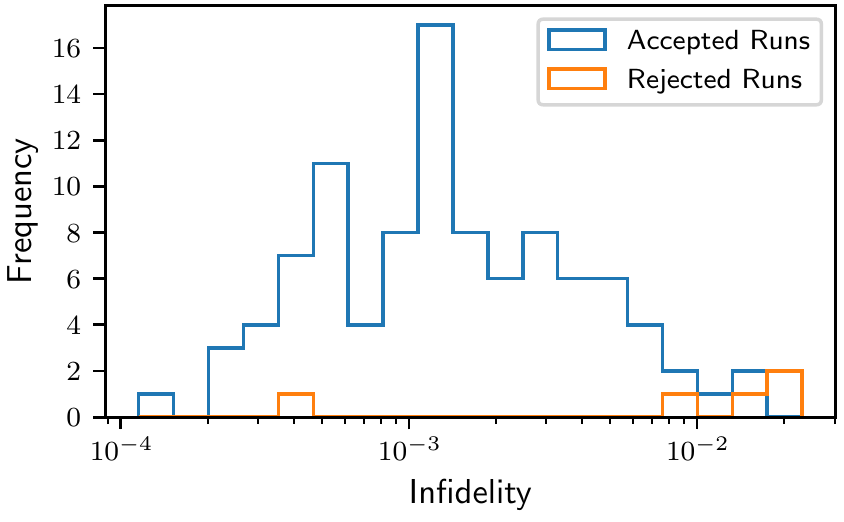}
    \caption{Histogram of gate infidelities due to imperfect calibration. Randomized benchmarking is used to compute a per-gate infidelity from the difference of the final parameter estimates of independent calibration runs compared to their mean value (Fig.~\ref{fig:endpoints_consistency}). We calculate a median infidelity of $1.3(1)\cdot 10^{-3}$ caused by imperfect calibration.}
    \label{fig:infidelities}
\end{figure}

\subsection{Confirmation Measurements}
\label{sec:confirm}

To use the calibration algorithm even with less than perfect success rate, we insert confirmation measurements after completing the procedure to detect and then reject outliers in the final gate parameters predicted by the algorithm. For rejected calibration runs the algorithm can then be repeated.

To detect outliers, we employ check sequences. We choose a sequence of 8 consecutive $\mathrm{MS}_0(\frac{\pi}{2})$ gates and a sequence of 6 gates with $\Delta\varphi_{target}=\pm\frac{\pi}{4}$ 
which both ideally return all populations to the ground state. The sequences are chosen because they are sensitive to Rabi frequency, phase difference miscalibrations respectively. 
We choose an acceptance threshold of at least 85 out of 100 measurements being in the target state. This threshold was chosen as a compromise between acceptance rate and the infidelity of the accepted runs, as more stringent thresholds start decreasing acceptance rate without significant improvements of either maximum or median infidelity of accepted runs.
We run repeated calibration runs including confirmation measurements and record the final parameter estimates produced by the calibration runs. $95(2)\%$ of runs pass the confirmation test (Fig.~\ref{fig:endpoints_consistency}). The distribution of accepted runs has standard deviations of $\sigma_{t_g}=1.9(1)\mathrm{\mu s},\; \sigma_{f_{cl}}=0.12(1)\cdot 2 \pi \mathrm{kHz}, \; \sigma_{f_{sb}}=0.21(2)\cdot 2 \pi \mathrm{kHz}, \; \sigma_{\phi}=0.025(3)\cdot \pi$. Translating these deviations to the gate parameters $\boldsymbol{\Theta}$ we find they are in agreement or below the thresholds set in Section \ref{sec:thresholds}.

We evaluate the estimated infidelity due to imperfect calibration of the accepted runs of the algorithm by simulating randomized cycle benchmarking using the difference between the final parameter estimates and the mean of all accepted calibration runs as miscalibration, and calculate the expected infidelity due to calibration error. We achieve a median calibration infidelity of $1.3(1)\cdot 10^{-3}$ (Fig.~\ref{fig:infidelities}). 
This value is lower than what was calculated for independent errors in section~\ref{sec:thresholds}, which we attribute to the correlations between the parameters partially compensating each other.
We can compare this to the fidelity of Bell states \cite{Jozsa1994} generated by repeated application of the MS gate on the same system, and find that the infidelity per gate is observed to be $4\cdot 10^{-3}$. While this state fidelity is not suited to characterize the effects of coherent errors, it does pose a limit on the achievable fidelity due to incoherent errors present in the system.
We expect laser phase noise and dephasing of the motional mode as the leading sources of decoherence. From this infidelity we conclude that our calibration routine can produce parameter estimates such that our entangling gates are not limited by calibration errors, but limited by the performance of the machine.

\section{Conclusion \& Outlook}
\label{sec:conOut}
In this work we proposed and implemented a procedure to use Bayesian parameter estimation to optimize the control parameters of an entangling two-qubit \MS gate. 
We described how the gate can be simulated as a function of a basic set of parameters, and how a particle filtering algorithm can be used to estimate them from measurements.
Two strategies for selecting the measurement settings are compared, and we show that a simple heuristic strategy can match the performance of a more complex variance minimization method.

We demonstrate that the algorithm can be used to calibrate a gate to a median residual infidelity due to parameter miscalibration of $1.3(1)\cdot 10^{-3}$ (Fig.~\ref{fig:infidelities}) in $1200 \pm 500$ experimental shots, achieving a greater than $30 \%$ speedup over manually iterating over 1D parameter scans (Appendix~\ref{appendix:manual_tuneup}). These results demonstrate the suitability of the described approach to reliably and consistently calibrate a set of control parameters to achieve a gate fidelity that is commensurate with the maximum fidelity set by incoherent processes in the quantum system. Our approach considerably eases the operational burden and can produce valuable time savings for the operating experimentalist. Furthermore, periodic re-calibration can be included within long experimental data-taking sequences, protecting against slow parameter drifts, and thus guaranteeing high quality output over considerably longer time.

The two-qubit gate optimisation routine we present here would benefit the quantum CCD ion trapped quantum computing architecture \cite{Kielpinski2002a}, where electric fields can be used to trap ions in multiple independent, spatially separate regions. By a combination of shuttling, splitting, and rotations, entangling operations can be performed between any 2 ions in the $N$ qubit register by isolating them in the appropriate `interaction' region \cite{Pino2021,Kaushal2020}. Once the gate is calibrated in the interaction region, owing due to all ions having identical atomic properties, the gate remains calibrated for the whole qubit register. The current implementation is also suited for experiments that generate 2-out-of-$N$ ion entanglement by selectively addressing ions in a 1D chain through their radial modes of motion \cite{Pogorelov2021}. In the presence of additional sources of miscalibration such as cross-talk and position dependent Stark shift, these schemes can easily be extended such that gates between each ion pair will have their own simulated interpolator. 
\par While the discussion here has focused on trapped ion implementations, we emphasize that this multidimensional Bayesian approach to gate calibration is not limited to just trapped ions and can be applied to any quantum system where a classical simulation of the entangling gate yields an accurate representation of the physical implementation. 

A limitation is that the classical information stored in the interpolator of the gate action will grow exponentially with the number of control parameters. We can thus ask what is the outlook for extending the scheme. The MS gate naturally scales to $N$ qubit entangling gates using the exact same set of control parameters. In this case the current scheme would work as long as the classical simulation remains tractable and can effectively describe other experimental imperfections such as magnetic field gradients or unequal Rabi frequencies on the individual ions. While this step might be time intensive for multiple qubits, it only needs to be performed once. When considering 2-out-of-$N$ entanglement generation on radial modes, modifications would be required in situations where many motional modes begin to participate in the MS interaction and complex amplitude or frequency modulation waveforms are required to disentangle the spins from these other modes \cite{Lu2019,Wright2019}. Extensions to include such optimal control techniques into the calibration protocol provide an avenue for further research. 

Another extension of the current scheme is the integration of more sophisticated control parameter tracking. Instead of periodically running the full calibration routine starting from an unbiased prior and $N_g=1$, one would retain the Bayesian prior from the previous calibration and seek to update it in as few measurements as possible to achieve further speed up. For the same amount of experimental shots, we could shorten the time between successive re-calibrations and would thus make the system even more robust against parameter drift as has been previously shown for single qubit systems \cite{Proctor2020,Ralph2011,Kimmel2015}. 

While the calibration is robust because runs that fail our convergence criterion are rejected, we continue to investigate the source of these occasional failures. We hypothesize that such events can be caused by large non-statistical deviations of the gate Hamiltonian due to transients in control fields. One method to characterize this would be to try an alternative automated calibration method, such as a machine learning approach, and compare the average achieved infidelity. This would allow us to characterize how sensitive competing methods are to such instabilities. Furthermore,  we  would  obtain  another  point  of comparison to benchmark the convergence speed against. In a similar vein, we could potentially enhance the success rate by further improving our classical simulation of the gate model by substituting the unitary predictions with a master equation description that includes measured noise sources such as heating and dephasing in our experimental setup. An interesting application of this open system extension would be to assess the suitability of our approach to calibrating experimental control systems that exhibit significant deviations from the ideal, unitary gate action.

\section{Acknowledgements}

We gratefully acknowledge support by the EU Quantum Technology Flagship grant AQTION under Grant Agreement number 820495, and by the US Army Research Office through Grant No. W911NF-14-1-010  and W911NF-21-1-0007.
We also acknowledge funding by the Austrian Science Fund (FWF), through the SFB BeyondC (FWF Project No. F7109), by the Austrian Research Promotion Agency (FFG) contract 872766, and by the IQI GmbH. MM acknowledges support by the ERC Starting Grant QNets Grant Number 804247. 

The research is also based upon work supported by the Office of the Director of National Intelligence (ODNI), Intelligence Advanced Research Projects Activity (IARPA), via the US Army Research Office Grant No. W911NF-16-1-0070. The views and conclusions contained herein are those of the authors and should not be interpreted as necessarily representing the official policies or endorsements, either expressed or implied, of the ODNI, IARPA, or the US Government. The US Government is authorized to reproduce and distribute reprints for Governmental purposes notwithstanding any copyright annotation thereon. Any opinions, findings, and conclusions or recommendations expressed in this material are those of the author(s) and do not necessarily reflect the view of the US Army Research Office.

\bibliography{bibliography.bib}

\appendix

\section{Comparison to different calibration methods}
\label{appendix:manual_tuneup}
One of the desiderata of an optimisation routine is having a fast convergence rate to the desired parameter accuracy. In other words we would like to minimize the number of times the experiment has to be queried since each measurement incurs by far the largest operational time overhead. Since in this publication we do not claim to have the optimal strategy, can we still claim that our Bayesian optimisation is fast relative to what might be considered a traditional or manual optimisation? In practice, each experimentalist might have their own slightly idiosyncratic way of optimising the gate parameters and will also have different amounts of prior knowledge about the parameter values. There is no set prescription that is followed by a majority consensus in the ion trap literature. However, reference \cite{Akerman2015} provides an algorithmic approach that iteratively performs one dimensional parameter scans to optimize the gate performance. We find this method closely mirrors what we would perform in our lab in the absence of the Bayesian optimisation routine and hence we will use it as a benchmark. 
In summary, the algorithm can be reduced to the following: 
\begin{enumerate}
    \item Roughly find the sideband detuning for desired gate time
    \item Scan the center line detuning at 2x gate time and maximize the population of $\ket{e,e}$
    \item 	Scan the sideband detuning at the presumed gate time and balance the populations $\ket{e,e}=\ket{g,g}$
    \item Scan the gate time and minimize the populations of $\ket{e,g}+\ket{g,e}$
    \item Repeat stepts 3 and 4
    \item Final repeat of step 3
    \item Repeat step 2
    \item Scan the Phase after two gates
\end{enumerate}

To quantize how many experimental queries (shots) we need to complete this algorithm at the desired precision, we make some very conservative boundary assumptions. We assume that the desired parameter can be extracted by fitting a Gaussian peak to each parameter scan and that the fit requires a minimum of 4 different parameter settings to constrain three variables - amplitude, standard deviation and centre point. It is this last quantity that needs to be extracted from such scans. We then require 50 shots per data point to achieve a $10^{-1}$ fractional error defined as the ratio of the standard error of Gaussian peak centre divided by the standard deviation of the Gaussian. This standard error is obtained from a weighted least squares fit to artificial data where the points lie equally spaced in an interval of 2 standard deviations on an ideal Gaussian curve while the weights are proportional to the inverse square of the binomial error from 50 trials. The fractional error decreases as $\propto 1/\sqrt{N}$ for N trials. The implication would be that, for example a scan such as in the top left of figure \ref{fig:parameter_scans} would be then defined to approximately 100 Hz which is a fair comparison to the threshold for the Bayesian optimisation. Given these assumptions, the above algorithm would require an absolute minimum of 1800 shots, which is already 50$\%$ more than the average performance of the Bayesian algorithm. In reality, typical scans will often use many more than 4 points as the prior knowledge of the parameters is lower than assumed in this analysis, scans sometimes need to iterated over repeatedly and the point spacing refined and the fitting is not as simple as assumed. In conclusion, the Bayesian algorithm outperforms a 1D manual parameter scan even granting for the most optimistic outcome and in typical operation is several times more efficient in the number of required shots. 

\end{document}